\documentstyle[12pt]{article}
\catcode`\@=11

\newcount\hour
\newcount\minute
\newtoks\amorpm
\hour=\time\divide\hour by 60
\minute=\time{\multiply\hour by 60 \global\advance\minute by-\hour}
\edef\standardtime{{\ifnum\hour<12 \global\amorpm={am}%
        \else\global\amorpm={pm}\advance\hour by-12 \fi
        \ifnum\hour=0 \hour=12 \fi
        \number\hour:\ifnum\minute<10
        0\fi\number\minute\the\amorpm}}
\edef\militarytime{\number\hour:\ifnum\minute<10
0\fi\number\minute}

\def\draftlabel#1{{\@bsphack\if@filesw {\let\thepage\relax
   \xdef\@gtempa{\write\@auxout{\string
      \newlabel{#1}{{\@currentlabel}{\thepage}}}}}\@gtempa
   \if@nobreak \ifvmode\nobreak\fi\fi\fi\@esphack}
        \gdef\@eqnlabel{#1}}
\def\@eqnlabel{}
\def\@vacuum{}
\def\marginnote#1{}
\def\draftmarginnote#1{\marginpar{\raggedright\scriptsize\tt#1}}
 \def \lc {light-cone\ }

\def\draft{
        \pagestyle{plain}
        \overfullrule=2pt
        \oddsidemargin -.5truein
        \def\@oddhead{\sl \phantom{\today\quad\militarytime} \hfil
        \smash{\Large\sl DRAFT} \hfil \today\quad\militarytime}
        \let\@evenhead\@oddhead
        \let\label=\draftlabel
        \let\marginnote=\draftmarginnote
        \def\ps@empty{\let\@mkboth\@gobbletwo
        \def\@oddfoot{\hfil \smash{\Large\sl DRAFT} \hfil}
        \let\@evenfoot\@oddhead}
        \def\@eqnnum{(\theequation)\rlap{\kern\marginparsep\tt\@eqnlabel}%
        \global\let\@eqnlabel\@vacuum}  }

\newcommand{\rf}[1]{(\ref{#1})}
\renewcommand{\theequation}{\thesection.\arabic{equation}}
\renewcommand{\thefootnote}{\fnsymbol{footnote}}

\newcommand{\newsection}{    % Numeration of eqs. is automatic
\setcounter{equation}{0}
\section}
\def\appendix#1{
  \addtocounter{section}{1}
  \setcounter{equation}{0}
  \renewcommand{\thesection}{\Alph{section}}
  \section*{Appendix \thesection\protect\indent \parbox[t]{11.15cm} {#1} }
  \addcontentsline{toc}{section}{Appendix \thesection\ \ \ #1}
  }

\def\apr{{{\rm A}^\prime}}
\def\bpr{{{\rm B}^\prime}}
\def\cpr{{{\rm C}^\prime}}

\def\epr{{{\rm E}^\prime}}

\def\sca{{\scriptscriptstyle{\cal  A}}}
\def\scb{{\scriptscriptstyle{\cal  B}}}

\def\aA{{\rm A}}
\def\aB{{\rm B}}
\def\aC{{\rm C}}
\def\aE{{\rm E}}

 \def\aA{{A}}
\def\aB{{B}}
\def\aC{{C}}
\def\aE{{E}}

\def\Csp{C^\prime}

\def\x'{\hbox{\'x}}
\def\y'{\hbox{\'y}}

\def\X'{\hbox{\'X}}

 \def \const {{\rm const}}
\def\alpr{i}
\def\bepr{j}
\def\gapr{k}

\def\vm{{\mu}}
\def\vn{{\nu}}

\def\ssmP{{\scriptscriptstyle P}}
\def\ssmK{{\scriptscriptstyle K}}
\def\ssmD{{\scriptscriptstyle D}}
\def\ssmQ{{\scriptscriptstyle Q}}
\def\ssmS{{\scriptscriptstyle S}}

\def\sfa{{\sf a}}
\def\sfb{{\sf b}}
\def\sfc{{\sf c}}
\def\dsfa{\dot{\sf a}}
\def\dsfb{\dot{\sf b}}
\def\dsfc{\dot{\sf c}}

\def \td {\tilde }

\def \gg {{\cal g}}

\def \foot {\footnote}
\def \bi{\bibitem}
\def \la {\label}

\def \ha {{1 \over 2}}
\def \ep{\epsilon}
\def \CC{{\cal C}}
\def \ov {\over}

\def\nline{\,\nabla\kern -0.7em\raise0.2ex\hbox{/}\,\,}
\def\yline{\,y\kern -0.47em /}
\def\aline{\,a\kern -0.49em /}
\def\parline{\,\partial\kern -0.55em /\,\,}

\def \t {\theta}
\def \s{\sigma}
\def \d {\partial}

\def\NPB#1(#2)#3{{\it Nucl. Phys.} {\bf B#1} (#2) #3}
\def\PRD#1(#2)#3{{\it Phys. Rev.} {\bf D#1} (#2) #3}
\def\PLB#1(#2)#3{{\it Phys. Lett.} {\bf B#1} (#2) #3}

\def\RMP#1(#2)#3{{\it Rev. Mod. Phys.} {\bf #1} (#2) #3}
\def\MPLA#1(#2)#3{{\it Mod. Phys. Lett.} {\bf A#1} (#2) #3}
\def\CQG#1(#2)#3{{\it Class. Quantum Grav.} {\bf #1} (#2) #3}
\def\AP#1(#2)#3{{\it Ann. Phys.} {\bf #1} (#2) #3}
\def\SJNP#1(#2)#3{{\it Sov. J. Nucl. Phys.} {\bf #1} (#2) #3}

\def \del{\partial}
\def \m {M}
\def \n {N}

\def\det{\hbox{det}}
\def\be{\begin{equation}}
\def\ee{\end{equation}}

\def \ci {\cite}
\def \g {\gamma}
\def \G {\Gamma}
\def \k {\kappa}

\def \L {{\cal L}}
\def \Tr {{\rm Tr}}
\def\apr{{A'}}

\begin{document}
%\draft
 %%%%%%%%%%%%%%%%%%%%%%%%%%%%%%%%%%%%%%%%%%%%%%%%%%%

\begin{titlepage}
\begin{flushright}
FIAN/TD/00-09
\\
OHSTPY-HEP-T-00-008
\\
hep-th/0007036\\
\end{flushright}
\vspace{.5cm}

\begin{center}
{\LARGE
%Light-cone  superstring in  AdS$_5 \times $S$^5$. I:
%$\k$-symmetry light cone  gauge
Superstring action  in AdS$_5 \times $S$^5$:\\[.1cm]
         $\k$-symmetry    light cone  gauge  }\\[.2cm]
\vspace{1.1cm}
{\large R.R. Metsaev${}^{{\rm a,b,}}$\footnote{\
E-mail: metsaev@lpi.ru, metsaev@pacific.mps.ohio-state.edu}
and A.A. Tseytlin${}^{{\rm a,}}$\footnote{\ Also at Blackett Laboratory,
Imperial College, London and   Lebedev Physics Institute, Moscow.\ \ \
 \ \ \ \ \ \ E-mail: tseytlin@mps.ohio-state.edu} }\\

\vspace{18pt}
 ${}^{{\rm a\ }}${\it
 Department of Physics,
The Ohio State University  \\
Columbus, OH 43210-1106, USA\\
}

\vspace{6pt}

${}^{{\rm b\ }}${\it
Department of Theoretical Physics, P.N. Lebedev Physical
Institute,\\ Leninsky prospect 53,  Moscow 117924, Russia
}
\end{center}

\vspace{2cm}

\begin{abstract}
%{\bf Abstract} \end{center}

As part of  program to quantize  superstrings in
$AdS_5 \times S^5$  in  a light-cone gauge
we  find the  explicit form of the corresponding   Green-Schwarz  action
 in the fermionic  light-cone  $\k$-symmetry gauge.
 The resulting action contains terms quadratic and quartic in fermions.
 In  flat space limit   it   reduces to  standard
 \lc GS  action, while    for  $\alpha'\to 0$  it
has correct  $AdS_5\times S^5$  \lc superparticle limit.
We  discuss  fixing the bosonic \lc  gauge
and   reformulation of the  action  in terms
of 2-d Dirac spinors.

\end{abstract}

\end{titlepage}
\setcounter{page}{1}
\renewcommand{\thefootnote}{\arabic{footnote}}
\setcounter{footnote}{0}

\def \adss {$AdS_5 \times S^5$\ }
\def \N {{\cal N}}
\def \lc {light-cone\ }
\def \ta { \tau}
\def \s { \sigma }
\def  \gg  { {\rm g}}
\def \sg {\sqrt {g }}
\def \te {\theta}
\def \vp {\varphi}
\def \gij {g_{ab}}
\def \xp {x^+}
\def \xm {x^-}
\def \p {\phi}
\def \vt {\theta}
\def \bx {\bar x} \def \a { \alpha}
\def \r {\rho}
\def \fourth {{1 \ov 4}}
\def \DD {{\cal D}}
\def \half {{1 \ov 2}}
\def \inv {^{-1}}
\def \ri {{\rm i}}
\def \D {{\cal D}}
\def \DD {{\rm D}}
\def \vr {\varrho}
 \def \diag {{\rm diag}} \def \td { \tilde }
\def \tta {\td \eta}
\def \cA {{\cal A}}
\def \cB   {{\cal B}}
\def \na {\nabla}

%%%%%%%%%%%%%%%%%%%%%%%%%%%%%%%%%%%%%%%%%%%
\newsection{Introduction and Summary}
%%%%%%%%%%%%%%%%%%%%%%%%%%%%%%

%%%%%%%%%%%%%%%%%%%%%%%%%%%%%%%%%%%%%%%
\subsection{Motivations for light-cone gauge approach}
%%%%%%%%%%%%%%%%%%%%%%%%%%%%%%%%%%%%%%%%%%%%%%%

The two maximally supersymmetric  backgrounds of  type IIB superstring
theory are  flat  Minkowski space $R^{1,9}$ and \adss.
The  manifestly supersymmetric
superstring action in  flat space -- the  Green-Schwarz (GS) action --
 is well  known \ci{GS},
and  its \adss analog  was constructed in \ci{MT}\
(see also \ci{KRR,MMT}).

 Progress in understanding  AdS/CFT duality \ci{mald}, i.e.
 in  solving  (the large $N$) supersymmetric $\N=4$ YM  theory in
 terms of  (first-quantized) superstring in \adss
 depends on   developing  its    GS  description
and making  it  more  practical.
 Some  advances in this direction  with application
 to ``long" strings ending at the boundary of
 $AdS_5$  were  discussed in  \ci{KT,Forste,DGT}.

While  the NSR string   action in  curved NS-NS  backgrounds
  has well-defined  kinetic terms and is at most quartic in
fermions, the GS action
in curved \adss background with R-R flux
  looks, in general, very non-linear
  \ci{MT,KRR,MMT}.
Its fermion structure
 simplifies in some special $\k$-symmetry gauges
  \ci{pessan,karl,kalram,KT},
  but, like in flat space,
   one may  still face the question of dependence
of the fermion kinetic term on a choice of bosonic
 string
background, i.e. of  its potential
degeneracy    \ci{KT}.

String configurations in \adss include ``short" closed strings
and ``long"  stretched strings  that may  end at the boundary.
The GS action is  well-suited  for description of small
  fluctuations near long string
backgrounds (for which  fermion kinetic term is well-defined).
 However,   to be able to determine  the fundamental
closed  string spectrum in \adss one is to learn how to quantize
the  \adss string action in the ``short string'' sector, i.e.
 without explicitly
 expanding near a  non-trivial  bosonic string
 configuration.

It is well-known how this is  achieved  for the
flat space GS action --
 by choosing a \lc gauge \ci{GSlc,GS}.
The superstring \lc gauge  fixing consists of the two  steps:

(I) fermionic \lc gauge choice,
i.e. fixing the $\k$-symmetry by  $\G^+ \theta^I=0$

(II) bosonic \lc gauge choice, i.e.
using   the conformal gauge\foot{We  use Minkowski
signature 2-d world sheet  metric $g_{\vm\vn}$ with
$g\equiv - \det g_{\vm\vn}$.}
$\sqrt {g}  g^{\vm\vn} =\eta^{\vm\vn}$
and fixing  the  residual conformal diffeomorphism symmetry
by $x^+(\ta,\s)  =  p^+ \tau$.

Fixing the  fermionic \lc gauge
  already produces a substantial
simplification of the flat-space
GS action: it   becomes  quadratic in $\theta$.
Choosing   the bosonic \lc
gauge, i.e.  using   an   explicit choice of $x^+$,
 may not always   be necessary
(cf. \ci{carlip,kalmor}), but it makes   derivation of the
physical string spectrum straightforward.

Our eventual  aim   is to develop a systematic \lc  gauge framework  for
 the GS  strings in \adss.
In this paper   we  shall concentrate on the first and   crucial
step of fixing the fermionic  \lc gauge, i.e. imposing
 an analog of $\G^+ \theta^I=0$ condition.\foot{A previous work in this direction
was reported in \ci{pes}, but the  $\k$-symmetry
\lc gauge used there is different from ours,
and we do not understand the relation of the  action presented in
\ci{pes}\ to our \lc gauge fixed  action.}
The  idea is to  get  a  simple gauge-fixed form
of the   action where the non-degeneracy of the kinetic term
for the  fermions  will not
 depend on a choice of a specific  string background
  in transverse directions,
  i.e., like in flat space, the fermion kinetic term will
  have the structure  $ \d \xp \bar \theta \d \theta$.

There are  other  motivations
for   studying   \adss strings   in  \lc gauge:

 (i) One of the prime goals is  to
 clarify the relation between the  string theory  and
  $\N=4$ SYM theory at the boundary.
 The SYM  theory does not admit a
 manifestly $\N=4$  supersymmetric  Lorentz-covariant description,
 but has a  simple superspace description in the
 \lc gauge $A^+=0$
 \ci{brink}. It is  based
 on a single  chiral superfield
 $\Phi(x, \te) = A(x) + \theta^i \psi_i(x) + ...$   where
 $A= A_1 + \ri A_2$ represents the  transverse
 components of the gauge field and $\psi_i$  its fermionic partner
 which transforms under the fundamental representation of R-symmetry group
 $SU(4)$.
 In addition to the  standard \lc  supersymmetry
 (shifts of  $\theta$),
 the \lc superspace SYM  action $S[\Phi]$
  has  also a non-linear superconformal symmetry.
This suggests that it may be possible   to formulate the bulk
string theory  in a way which is  naturally related to the \lc  form
of the boundary SYM theory. In particular,  it may be useful
to  split the  corresponding   fermionic string   coordinates
into  the   two  parts  with  manifest $SU(4)\simeq SO(6)$
transformation properties
which  will be the counterparts of the linearly realized
Poincar\'e supersymmetry  supercharges
and  the  nonlinearly realized conformal supersymmetry  supercharges
of the  SYM  theory.

 (ii) As was shown in \ci{metsi,metsii,metsiii},
field theories  in AdS space  (in particular, IIB supergravity)
admit a   simple  \lc description. There exists
a \lc   action for a superparticle in \adss
which was used to formulate AdS/CFT correspondence in  the \lc gauge.
This  suggests that the full superstring theory in \adss should also have
a  natural \lc gauge  formulation, which should be useful
in the context of   the AdS/CFT correspondence.

  %%%%%%%%%%%%%%%%%%%%%%%%%%%%%%%%%%%%%%%%%%%%%%%%%%%%%%%%%%%%%%%%
 \subsection{Structure of the light-cone gauge  string action}
 %%%%%%%%%%%%%%%%%%%%%%%%%%%%%%%%%%%%%%%%%%%%%%%%%%%%%%%%%%%%%%%%%

Our  fermionic  $\k$-symmetry
\lc gauge (which is  be different from the
naive $\G^+ \theta^I =0$ but is  related  to it in the flat space limit)
will reduce the 32 fermionic coordinates $\theta^I_\a$ (two left
Majorana-Weyl 10-d spinors)
to 16 physical Grassmann variables:
``linear"  $\te^i$  and   ``nonlinear" $\eta^i$
and their hermitian conjugates $\te_i$  and $\eta_i$ ($i=1,2,3,4$),
which  transform
according to the fundamental representations of $SU(4)$.
The superconformal algebra $psu(2,2|4)$  dictates that
these variables   should  be  related to  the  Poincar\'e
and  the  conformal
supersymmetry   in the \lc gauge description of the
    boundary theory. The action and symmetry generators will
     have simple (quadratic)
 dependence on
$\te^i$, but complicated (quartic)
dependence on $\eta^i$.\footnote{These coordinates
are direct counterparts of the Grassmann coordinates
in the \lc
action  for a superparticle in \adss\  in  \ci{metsii,metsiii}.
}

We shall split  the 10  bosonic coordinates of \adss as  follows:
the 4 isometric  coordinates  along the boundary directions  will be
\begin{equation}\label{lcbas}
x^a= (x^+,x^-,x, \bx)\,,
\qquad
x^\pm\equiv \frac{1}{\sqrt{2}}(x^3\pm x^0)\,,
\qquad
x,\bx = { 1 \ov \sqrt 2} (x^1 \pm  {\rm i} x^2)\,,
%  \qquad \bar x={ 1 \ov \sqrt 2} (x^1 - {\rm i} x^2)\,,
\end{equation}
the radial direction of $AdS_5$ will be  $\p$,
and the  $S^5$  coordinates will be denoted as   $y^\apr$
($\apr=1,2,3,4,5$).

Choosing a  light-cone  gauge in the parametrization  of the  supercoset
 $PSU(2,2|4)/$ $[SO(4,1)$ $\times SO(5)]$
 described below,
the \adss superstring
 Lagrangian  of  \ci{MT} can be written as\footnote{The
 light-cone  gauge action  can be  found  in two related forms. One of them
corresponds to the Wess-Zumino type gauge in 10-d superspace
while another is based on the  Killing gauge (see \ci{KRR,karl}).
 These ``gauges"    (better to be called ``parametrizations")
  do not reduce the number of fermionic degrees of
freedom but only specialize a  choice  of fermionic coordinates.
The action given in this Section 
  corresponds to the  WZ parametrization,
while the action   in the Killing    parametrization 
will be discussed in  Section 6.
 }
 \be
{\cal L}={\cal L}_{B}+ \L^{(2)}_{F} + \L^{(4)}_{F} \ .
%  \L^{(2)}_{F} = \L_{kin} + {\L}_{WZ} \ .
\la{oom}
\ee
Here $\L_B = -\ha \sqrt{g} g^{\vm\vn}  G_{\m\n}(X)
\del_\vm X^\m \del_\vn  X^\n$  is the
standard bosonic sigma model with  \adss as target
space,\foot{Our  index notation differs
from \ci{MT}: here we use  $\vm,\vn=0,1$ for 2-d indices,
$i,j$ for $SU(4)$ indices, $A=0,1,...,4$ for $AdS_5$ and
$\apr=1,..,5$   for $S^5$ tangent space indices
(repeated indices are contracted with flat metric).
We use   $\epsilon^{01}=1$.
 }
\be
\L_B = - \sg g^{\vm\vn} \bigg[
e^{2\phi}(\partial_\vm x^+ \partial_\vn x ^-
+ \partial_\vm x\partial_\vn\bar{x})
+ \frac{1}{2}\partial_\vm \phi \d_\vn \p
 + \ha e_\vm^\apr e_\vn^\apr \bigg] \ .
\la{bos}
\ee
$e_\vm^\apr$ is the projection of the vielbein of $S^5$  which
 in the special parametrization we will be  using  is given by
\be
 e_\vm^\apr   = -\frac{\rm i}{2}
\Tr(\gamma^\apr \partial_\vm UU^{-1})\ , \ \ \
\  \ \ \  \ U^i{}_j\equiv (e^y)^i{}_j\ , \ \
\ \
U^\dagger U=I\ , \la{uuu}
\ee
where $\Tr$ is over $i,j$.
The matrix  $U\in SU(4)$ depends on 5 independent coordinates
$y^\apr$
\be
y^i{}_j\equiv \frac{{\rm i}}{2}y^\apr (\gamma^\apr)^ i{}_j \ ,
\ \ \ \ \ \
(y^i{}_j)^* =-y^j{}_i\,,\  \ \ \
\
y^i{}_i=0\,.
\la{dere}
\ee
and   $\gamma^\apr$ are $SO(5)$ Dirac matrices.
$  \L^{(2)}_{F}$ is    the quadratic part of the  fermionic    action
\begin{equation}\label{ff}
{\cal L}^{(2)}_{F}
=e^{2\phi}\partial_\vm x^+\Bigl[\frac{{\rm i}}{2} \sqrt{g}g^{\vm\vn}
(-\theta_i {\cal D}_\vn\theta^i -\eta_i {\cal D}_\vn \eta^i
+{\rm i}\eta_i e_\vn^i{}_j\eta^j)
+\epsilon^{\vm\vn}\eta^i C_{ij}^\prime
({\cal D}_\vn\theta^j 
-{\rm i}\sqrt{2}e^\phi\eta^j\partial_\nu x)\Bigr]
+h.c.
\end{equation}
The  P-odd $\ep^{\mu\nu}$ dependent  term in \rf{ff} came from the
WZ term in the original supercoset GS action \ci{MT}.

Here  we used the following notation
\begin{equation}\label{covder}
{\cal D}\theta^i=d\theta^i-\Omega^i{}_j\theta^j\,,
\qquad
{\cal D}\theta_i=d\theta_i+\theta_j\Omega^j{}_i\,,
\qquad
e^i{}_j \equiv (\gamma^\apr)^i{}_j e^\apr \ ,
\end{equation}
and $  {\cal D} = {\cal D}_\mu d \s^\mu, \  e^i{}_j = e_\mu^i{}_j d\s^\mu $
where $\s^\vm=(\tau,\sigma)$ are 2-d coordinates.
${\cal D}$ is the generalized spinor derivative on $S^5$.
It has  the general
representation
${\cal D}=d+\Omega^i{}_j J^j{}_i$ and satisfies the relation
${\cal D}^2=0$.
$\Omega^i{}_j$ is given by
\begin{equation}\label{gencon}
\Omega= dUU^{-1}\,,
\qquad
d\Omega-\Omega\wedge \Omega=0   \ ,
\end{equation}
and can be written  in terms of the $S^5$  spin  connection
$\omega^{\apr\bpr}$ and the  5-bein $e^\apr$ as follows
 \begin{equation}\label{gencon1}
\Omega^i{}_j=-\frac{1}{4}(\gamma^{\apr\bpr})^i{}_j\omega^{\apr\bpr}
+\frac{\rm i}{2}(\gamma^\apr)^i{}_je^\apr   \ .
 \end{equation}
$\Csp_{ij}$ is the constant charge conjugation matrix
of the $SO(5)$ Dirac matrix algebra: ${\Csp}^\dagger \Csp = I,
{\Csp}^T =- \Csp$.
The hermitean conjugation rules are:
$\theta_i^\dagger=\theta^i$, $\eta_i^\dagger=\eta^i$.

The quartic fermionic term in \rf{oom}
depends  only  $\eta$  but not on  $\theta$
\be
\L^{(4)}_{F} = { 1 \ov 2} \sg g^{\vm\vn}
 e^{4\phi}\partial_\vm x^+ \d_\vn x^+  \bigg[ (\eta^i\eta_i)^2  -
(\eta_i \gamma^{\apr i}{}_j \eta^j)^2 \bigg] \ .
\la{qua}
\ee

 %%%%%%%%%%%%%%%%%%%%%%%%%%%%%%%%%%%%%%%%%%%%%%%%%%%%%%%%%%%%%%%%
 \subsection{Some properties  of the  action}
 %%%%%%%%%%%%%%%%%%%%%%%%%%%%%%%%%%%%%%%%%%%%%%%%%%%%%%%%%%%%%%%%%

The  action (\ref{oom}),(\ref{bos}),(\ref{ff}),(\ref{qua})  has several
important properties:

(a) The  dependence on $x^-$ is only  linear -- through
the  bosonic $\del x^+ \del x^-$ term in \rf{bos}.

(b) The bosonic factor in the    fermion
kinetic  term  is simply $e^{2 \p} \d \xp$.
It is the crucial property of this \lc $\k$-symmetry gauge
fixed form of the action  that the fermion  kinetic
term involves  the derivative of
only  {\it one} space-time direction
-- $x^+$, i.e. its (non)degeneracy
does not depend on  transverse
string  profile.\foot{The action thus has  similar  structure
 to that of  the \lc gauge  action for the GS string in curved magnetic R-R
 background  constructed in \ci{RT}.}

(c) The fact that the action  has  only  quadratic and quartic
fermionic terms  has to do with special symmetries
of the \adss background (covariantly
constant curvature and 5-form  field strength).
The presence of the  $\eta^4$ term \rf{qua}  reflects
the curvature of the background.\foot{Note that
the  light-cone gauge
GS action in a curved space of the form $R^{1,1} \times M^8$
with generic  NS-NS and R-R backgrounds  \ci{FRT}
(reconstructed
from the \lc flat space  GS vertex operators \ci{GGSS})
contains,  in general,  higher than quartic fermionic terms,
multiplied by higher derivatives of the background fields.
This     \lc  GS
action has  quartic fermionic term \ci{HW,FRT}
involving the curvature tensor \
$R_{....} \d \xp  \d \xp  (\bar \theta\G^{-..} \vt )(\bar \theta  \G^{-..} \theta)
\sim
R_{....} (p^+)^2  (\bar \theta \G^{-..} \vt)  ( \bar \vt
\G^{-..} \vt) $ which is similar to the one present in the
NSR string action
(i.e. in  the standard 2-d supersymmetric sigma model).}
As follows from the discussion in \ci{MT},
the  `extra' terms in (\ref{ff}) like $\eta_i e^i{}_j\eta^j$ and
$\eta \Csp \eta \partial x$ should have the interpretation of the
couplings to the R-R  5-form  background.\foot{The part of the action in
\ci{MT} quadratic in $\vt^I$ is a direct generalization of the quadratic
term in the flat-space GS action (before $\k$ symmetry  gauge fixing) $
S^{(2)}_{F} = {{\rm i }\ov 2 \pi \a'} \int d^2 \s ( \sqrt{g}
g^{\vm\vn}\delta^{IJ} - \ep^{\vm\vn} s^{IJ} ) \bar \vt^I \rho_\vm D_\vn
\vt^J .  $ Here $\r_\vm$ are projections of the 10-d Dirac matrices, $
\r_\vm \equiv \G_{\hat m} E^{\hat m}_\m \del_\vm X^\m = ( \G_A E^A_\m
+\G_\apr E^\apr_\m ) \del_\vm X^\m \ , $ and $E^{\hat m}_\m$ is the
vielbein of the 10-d target space metric.
%$ G_{ \mu \nu } = E^{\hat a }_\m  E^{\hat b}_\n %\eta_{\hat a \hat b}.$
The covariant derivative $D_\mu$ is the projection of the
10-d derivative
$D_{ M}=\del_{ M}
+ \fourth\omega^{\hat m\hat n}_{ M} \Gamma_{\hat m\hat n}
 - { 1 \ov 8 \cdot 5!}
 \G^{ \m_1...\m_5} \G_{ \m}\ e^\Phi F_{ \m_1... \m_5}$
%($\Omega^{\hat m\hat n}_\m$ is the spin connection and
%$F_{ \m_1... \m_5}$ the RR 5-form potential)
which appears in the Killing spinor equation of type
IIB supergravity. It has the following explicit form
$
D_\mu\t^I \equiv \left(\delta^{IJ} {\rm D}_\mu
- {\ri\ov 2 } \epsilon^{IJ} \tilde\r_\mu \right) \vt^J,$
\ \  $
{\rm D}_\mu = \del_\mu
+\fourth \del_\mu  x^\m \omega^{\hat m\hat n}_\m\Gamma_{\hat m\hat
n}, $    where the term with
 $\tilde\r_\mu \equiv \left(\G_A E^A_\m +\ri\G_\apr E^\apr_\m\right)
\del_\mu  X^\m $   originates from the coupling
 to the R-R 5-form field strength.}

(d)  The  gauge we considered treats      the
$AdS_5$ and $S^5$ factors asymmetrically. In particular,
the action  contains only $SO(5)$ but not $SO(4,1)$
gamma matrices, and $\te_i$ and $\eta_i$ are  {not}
spinors  under $SO(4,1)$.\foot{ $\theta$ and $\eta$ are not
scalars with respect to $SO(4,1)$.
Combined together with fermions eliminated  by $\kappa$-symmetry gauge
 they transform in
spinor representation
 of $SO(4,1)\times SO(5)$. But after gauge fixing which is based on
$\gamma$ matrices from $AdS_5$ part ($\gamma^+\theta=0$),  the $SO(4,1)$
group, with the exception of its $SO(2)$ subgroup
generated by  $J^{x\bar{x}}$
\ci{metsi} (which is part of little group  for the
$AdS_5$ case)  becomes  realized nonlinearly.
 Thus (modulo subtleties of nonlinear realisation of
$su(4)$ on bosons) the algebra
$so(2)\oplus su(4)$ is  a counterpart of the algebra
$so(8)$ in  flat case.}

(e) The  \adss superstring action depends on  two  parameters:
the scale   (equal radii)  $R$ of $AdS_5\times S^5$ and the inverse string tension
or $\alpha'$.
Restoring the dependence on    $R$
  set equal to 1    in \rf{oom}
 one finds that in the flat space limit    $R\to \infty$
 the quartic term \rf{qua} goes away, while
  the kinetic term \rf{ff}
 reduces to the standard  one  with ${\cal D}_\mu \to \d_\mu$.
 The resulting action is equivalent to the flat space
 \lc GS action \ci{GS} after representing  each of the
 two $SO(8)$ spinors in terms of the two  $SU(4)$ spinors.
 The action takes `diagonal' form in terms of the
 combinations $ \psi_{1,2}^i$ of our two fermionic variables (see
 (\ref{combin}) below).

 (f) For $\alpha'\to 0$
the action has the correct particle limit, i.e. it
reduces to the light-cone gauge superparticle action in
$AdS_5\times S^5$ \cite{metsii}.

(g) A special feature of this  action is
that  $SU(4)\simeq SO(6)$ symmetry  is realized linearly
on fermions,
but not on bosons, i.e. is not manifest. This is a
consequence of the factor $SO(4,1) \times SO(5)$
 in the underlying
supercoset  $PSU(2,2|4)/[SO(4,1) \times SO(5)]$
being   purely bosonic.  The
$S^5= SO(6)/SO(5)$ part of the bosonic action
can be represented as a special case of the 2-d
$G/H$  coset sigma model
$ L = \Tr ( \d_\mu {\cal U} {\cal U} \inv + A_\mu )^2$,  \  ${\cal U}  \in G=SO(6) $, \
with  the 2-d gauge field
$A_\mu$ being in the algebra of $H= SO(5)$.
This  action does not have
manifest $SO(6)$ symmetry after $A_\mu$ is integrated out
and  ${\cal U}$ is restricted to belong to the
coset as a gauge choice.

(h) The action is symmetric under shifting $\theta\rightarrow
\theta+\epsilon$
  supplemented by an appropriate transformation of $x^-$.
Here $\epsilon$ is a Killing spinor on $S^5$,
 satisfying  the
equation ${\cal D}\epsilon^i=0$. It is this symmetry that is
responsible for the fact that
the theory is linear in $\theta$, i.e.
that there is no quartic interactions in $\theta$.

 %%%%%%%%%%%%%%%%%%%%%%%%%%%%%%%%%%%%%%%%%%%%%%%%%%%%%%%%%%%% \bigskip
 \subsection{ Bosonic \lc gauge fixing and  elimination of   $x^+$ }
%%%%%%%%%%%%%%%%%%%%%%%%%%%%%%%%%%%%%%%%%%%%%%%%%%%%

To proceed further to quantization of the theory
one would like, as in the flat   case,
 to eliminate the $\del x^+ $-factors from the fermion kinetic
 terms  in \rf{ff}.
 In flat space this  was  possible by choosing the  bosonic \lc
 gauge.    In the  BDHP formulation \ci{bdh,poly}
 which we  are using this may be  done by fixing the conformal gauge
   \be
 \g^{\vm\vn}   = \eta^{\vm\vn}\ , \ \ \ \ \
  \g^{\vm\vn}\equiv \sg g^{\vm\vn} \ ,  \ \ \ \  \det  \g^{\vm\vn}  =-1
   \ , \la{coonf} \ee
     and then noting
   that since the resulting equation
   $\del^2 x^+=0$ has the general solution
   $x^+ (\tau,\s) = f( \tau-\s) + h( \tau+ \s)$
   one can fix the residual conformal diffeomorphism symmetry
   on the plane by choosing $x^+(\tau,\sigma)  = p^+ \tau$.
 An  alternative (equivalent) approach
  is   to use  the original  GGRT \ci{ggrt}
 formulation  based on writing the Nambu action in the canonical  first
  order form (with constraints added  with Lagrange multipliers)
  and fixing the
  diffeomorphisms  by  2  conditions -- on  one coordinate and one canonical
  momentum:
  $x^+ = p^+\tau, \ P^+  =\const$.\foot{Yet another approach is to fix
  $g_{--} =0, \ x^+ = h(\tau,\s)$ where $h$ is determined by external sources
  \ci{gers}.
  For a discussion of various ways
  of fixing the \lc gauge   in the case of flat target space   and their
  relations see, e.g., \ci{smitze}.}

The  first  approach  based on the conformal gauge
 does not in general  apply in curved spaces
with null Killing vectors which are not of the direct product
form $R^{1,1} \times M^8$ (the gauge conditions will not
in general be consistent with classical equations of motion).
It does apply, however,
if the null Killing vector is covariantly constant \ci{hors}.
There is no need, in principle, to insist on fixing
 the standard conformal gauge   \rf{coonf}.
Instead, one may fix the diffeomorphism gauge
  by  imposing  the two conditions
 $\g^{00}=-1, \   x^+ = p^+ \tau$. This choice   is consistent
  provided
the background metric satisfies
$G_{+-}=1, G_{--}=G_{-i}=0, \del_- G_{MN} =0$
 \ci{rudd}.  This approach is  essentially equivalent to
 the GGRT approach applied to the curved space case.

The above conditions do not apply in the $AdS$  case:
the  null Killing vectors  are  not covariantly constant
 and $G_{+-} = e^{2\phi} \not=1$.\footnote{In fact, there is no
 globally well-defined null Killing  vector  in $AdS$ space 
 as its  norm proportional to $e^{2\phi}$
 vanishes at the horizon $\phi=-\infty$
 (this point and a  possibility to 
  fix a global diffeomorphism gauge 
for  $AdS$ string  was discussed in \ci{roz}).
In this paper we shall use  a formal approach to this issue:
since  the boundary SYM theory 
in $R^{1,3}$  space has a well-defined light-cone description,
it should be possible to fix 
 some analog of a \lc gauge  for the
dual string as well (assuming it is defined on the 
Poincare patch  of the $AdS$ space).
A potential  problem of that approach  which will be 
 reflected in the
degeneracy of the resulting \lc gauge fixed action 
near the horizon region should then be addressed 
at a later stage.}
 It is easy to see, however,  that
  a slight modification    of the above  conditions  on $\g^{00}, \ x^+$
 represents  a consistent  gauge choice
  \be
 e^{2\phi}\g^{00}=-1\ , \   \ \ \ \ \ \ \ x^+ = p^+ \tau  \ .
 \la{our}
 \ee
 Indeed,  the  equation for $x^+$
 \be
 \del_\mu ( e^{2 \p} \sg g^{\vm\vn} \del_\vn x^+) =0 \
\la{con}
\ee
 is then satisfied.
 The coordinate space BDHP  approach based on \rf{our} is equivalent
 to the phase space GGRT approach   based on  fixing the diffeomorhisms by
   $x^+ = p^+\tau, \ P^+  =\const$.   The possibility
   to fix the  \lc gauge  for the bosonic string in $AdS$  space
   using the latter GGRT approach was originally suggested
   by Thorn \ci{thornn}.

 A complication  in the case of fixing  the diffeomorphisms
 by the conditions on $\g^{00}$ and $x^+$
  (or on $P^+$ and $x^+$ in the phase space approach)
compared to the cases where one can fix  the   2-d metric completely
by choosing the
 conformal gauge
  is that here one is still to integrate over
 the remaining independent component of the 2-d metric (or $\g^{01}$)
 and to solve the resulting constraint.
One may try  to avoid this   by
  fixing instead   a modification  of the  conformal gauge \rf{coonf}
  suggested by Polyakov  \ci{pol}
  \be
 \g^{\vm\vn} = \diag(- e^{-2\p}, e^{2\p})\ ,
 \la{cho}
  \ee
such that  \rf{con}   still  has   $x^+ =p^+ \tau$ as its
solution.   This is just a particular classical
solution,  and it may seem that  in contrast to the 
flat space case
here one is unable to argue
that     $x^+ =p^+ \tau$  represents a gauge
 fixing  condition for  some  residual
symmetry.  However, this ansatz  may indeed be   justified 
a posteriori as  being the outcome of  a  systematic procedure 
based on fixing $x^+$ and one condition on 2-d metric like
\rf{our}   and then integrating over $x^-$
(assuming it has no sources). 
%(see  below). 

% With this understanding it
%  may still be useful to develop  string perturbation theory
% in the gauge \rf{cho} by expanding near this special  classical
% `light-cone' solution     $x^+ =p^+ \tau$.

In this paper we shall  not discuss in detail 
the consequences of fixing the  bosonic
\lc gauge \rf{our}  in the  superstring action   \rf{oom}
(or  the equivalent  \lc gauge fixing
in the phase space GGRT approach \ci{ MTT})
and  follow a  simplified  approach  based
on  using  a particular classical solution.

 Let us  first not make any explicit gauge choice
 and consider the superstring path integral
  assuming   that there is no sources for $x^-$.
The linear dependence of the action \rf{oom},\rf{bos}
on $x^-$  allows  us  to  integrate over    $x^-$
explicitly.
This
 produces  the $\delta$-function constraint  imposing the equation of motion
 \rf{con}
 for $x^+$, which
  is formally solved by    setting
\be
\sg g^{\vm\vn} e^{2 \p} \del_\vn x^+ = \ep^{\vm\vn} \del_\vn  f
\ ,
\la{solv}
\ee
where $f(\tau,\s)$ is an {\it arbitrary}  function.
Since our action \rf{oom}  depends only on  $x^+$
only through $e^{2 \p} \del x^+ $,
we are then   able to integrate over $x^+$ as well,
eliminating it
in favor  of the function  $f$.
The  action will contain the fermionic terms
 \rf{ff},\rf{qua} with
  \be  e^{2 \p} \del_\mu x^+\ \ \to \ \
  f_\mu \equiv g_{\mu\nu }{ \ep^{\nu\lambda} \ov \sg } \del_\lambda  f\ .  \ee
The resulting fermion kinetic term is  then
non-degenerate (for a properly chosen  $f$), and may  be
interpreted as  an  action of 2-d fermions in curved 2-d geometry
 determined by $f $ and $g_{\mu\nu}$ (cf. \ci{DGT,sedr,wig}).

We  may then  simplify the action further by
   making a special choice of $f$   and
 fixing  a  diffeomorphism gauge on $g_{\mu\nu}$
 in a consistent way.
One possibility is to  choose  the gauge \rf{cho}
and $f\sim \s$   which implies according to \rf{solv} that
$x^+\sim  \tau$, i.e.\foot{ Note that the standard conformal   gauge
$\sg  g^{\vm\vn} = \diag(-1, 1)$  leads to inconsistency for generic
$\p$ if one insists on the simplest $f=\s$ choice.
 Consistency for generic $\p$  is achieved
 only if $f$ (and $x^+$) are nontrivial.
But then the structure of the resulting action is
complicated.}
 \be
 \la{choi}
 f= \s\ , \ \ \ \ \  x^+ = \tau\ , \ \ \ \ \ \
 \
 \sg g^{\vm\vn} = \diag(- e^{-2\p}, e^{2\p})\ .
 \ee

%%%%%%%%%%%%%%%%%%%%%%%%%%%%%%%%%%%%%%%%%%%%%%%%%%%%%%%%%%
 \subsection{``2-d  spinor"  form  of the  action  }
%%%%%%%%%%%%%%%%%%%%%%%%%%%%%%%%%%%%%%%%%%

Like in  the flat space case \ci{GS}
and in the ``long string" cases discussed
in  \ci{DGT}  the
resulting   action   can  then  be  put
into  the ``2-d spinor" form.  Indeed,
the   8+8  fermionic degrees of freedom  can be
organized into 4  Dirac 2-d spinors,
 defined  in  {\it curved} 2-d geometry.
Using  \rf{choi}  we can  write the kinetic term \rf{ff} as
\begin{equation}\label{ff2}
{\cal L}^{(2)}_{F}
=\frac{{\rm i}}{2}(\theta_i {\cal D}_0\theta^i + \eta_i {\cal D}_0 \eta^i
-{\rm i}\eta_i e_0^i{}_j\eta^j)
+ e^{2\phi}\eta^i C_{ij}^\prime
({\cal D}_1\theta^j - {\rm i}\sqrt{2}e^\phi\eta^j\partial_1 x)
+h.c.
\end{equation}
Introducing a  2-d zweibein  corresponding to the metric in \rf{choi}
\be
e_\vm^m = \diag(e^{2\phi}, 1)\,,
\qquad
g_{\vm\vn} = - e^0_\vm e^0_\vn + e^1_\vm e^1_\vn \ ,
\la{gtg}
\ee
we can put  \rf{ff2}  in the 2-d form as follows
\begin{equation}\label{ff3}
e^{-1}{\cal L}^{(2)}_{F}
=-\frac{\rm i}{2}\bar{\psi}\varrho^m e_m^\vm {\cal D}_\mu\psi
+\frac{\rm i}{2}\bar{\psi}\psi\partial_1\phi
-\frac{1}{\sqrt{2}}\bar{\psi}_ie_0^i{}_j\varrho^-\psi^j
+{\rm i}\sqrt{2}e^\phi (\psi^{i})^T \pi^-\Csp_{ij}\psi^j\partial_1\bar{x}
+h.c. \
\end{equation}
Here  $\varrho^m$  are 2-d Dirac matrices,
\be\varrho^0= i \s_2,
\quad
\varrho^1 = \s_1,
\quad
\varrho^3 =\varrho^0 \varrho^1= \s_3\,,
\quad
\varrho^\pm\equiv \frac{1}{\sqrt{2}}(\varrho^3\pm \varrho^0)\,,
\qquad
\pi^-\equiv \frac{1}{2} ({1-\varrho^1})\ ,
\la{dirr}\ee
  $\bar \psi_i= (\psi^i)^\dagger \varrho^0$,\
$\bar{\psi}\psi$ stands for $\bar{\psi}_i\psi^i$,
$\psi^T$ denotes the transposition of 2-d spinor  and
$\psi$'s are related to the  original (2-d scalar) fermionic  variables
$\theta$'s and $\eta$'s   by\foot{In our notation  \ \
${\rm i} \bar \psi \varrho^m  \na_m\psi
= - \ri \psi_1 ^\dagger (\na_0 - \na_1) \psi_1 -
\ri \psi_2 ^\dagger (\na_0 + \na_1) \psi_2$, \ \
$\na_m= e_m^\vm \del_\vm$.}
\begin{equation}\label{combin}
\psi^i= \left(
\begin{array}{c}
\psi^i_1\\
\psi^i_2
\end{array}\right)\,,
\qquad
\psi_1^i=\frac{1}{\sqrt{2}}[\theta^i-{\rm i}({\Csp}^{-1})^{ij}\eta_j]\,,
\qquad
\psi_2^i=\frac{1}{\sqrt{2}}[\theta^i+{\rm i}({\Csp}^{-1})^{ij}\eta_j]\,.
\end{equation}
The quartic interaction term \rf{qua}
 takes the  following form
\begin{equation}\label{qua2}
e^{-1}\L^{(4)}_{F} =
\frac{1}{4}\Bigl[(\bar{\psi}_i \gamma^{\apr i}{}_j\varrho^-\psi^j)^2
-(\bar{\psi}_i\varrho^-\psi^i)^2\Bigr] \ .
\end{equation}
 The total action is thus a  kind of
  $G/H$ bosonic sigma model coupled to
  a  Thirring-type  2-d
fermionic model  in  curved 2-d geometry  \rf{gtg}
(determined by the profile of the
radial function of the $AdS$ space), and
coupled to  some 2-d vector fields.
The interactions  are such that they  ensure the quantum
2-d conformal invariance
of the total model \ci{MT}.

Properties of the resulting action
and whether it can be put into  simpler and useful form
remain to be studied.
 It is clear of course that the action
 has  a rather  complicated  structure   and is
 not solvable in terms of free fields
 in any obvious  way. A hope is that the
 \lc form of the action we  have found
 (or its first order phase space  analog)
 may suggest  a choice of more  adequate variables
 which may allow further progress.

%%%%%%%%%%%%%%%%%%%%%%%%%%%%%%%%%%%%%%%%%%%%%%%%%%%%%%%%%%
We finish  this discussion with few remarks:

(i)
The mass term $ \bar{\psi}\psi\partial_1\phi$
in (\ref{ff3}) is similar to the one in \cite{DGT}
(where the  background string configuration was
non-constant only in the radial $\phi$ direction)
and  has its origin  in the
 $\epsilon^{\mu\nu }  e^{2\p} \del_\mu x^+ \del_\nu \phi
\eta^i\Csp_{ij} \theta^j$ term
appearing after $\eta \leftrightarrow \theta$ symmetrization of  the
$\epsilon^{\vm\vn }$ term in \rf{ff}
 (its `non-covariance' is thus a consequence of the choice
$x^+=\tau$).

(ii) The action is symmetric under shifting \
$
\psi^i \rightarrow \psi^i + \varrho^- \epsilon^i     , $
where $\epsilon^i$ is the 2-d  Killing  spinor. This
symmetry reflects the fact that our original action is symmetric under
shifting  $\theta^i$ by a Killing spinor on $S^5$.

(iii)  The 2-d Lorentz invariance
 is  preserved  by the fermionic \lc
gauge (original GS fermions   $\vt$  are 2-d scalars)
 but is broken by our special choice
of the  bosonic gauge \rf{choi}.
The special form  of $g_{\vm\vn}$  in \rf{choi}
implies  ``non-covariant"  dependence on $\p$
in the  bosonic part of the action:
  the action
\rf{bos} for  the 3 fields $\p,x,\bx$
and  the  5-sphere    coordinates  $y^\apr$
has the form
$$
\L_B =  \d_0 x \d_0 \bx  - e^{4\p} \d_1 x \d_1 \bx
+ \ha   e^{-2\p} \d_0 \p  \d_0 \p   - \ha e^{2\p} \d_1 \p  \d_1 \p
$$
\be
+\  \ha G_{\sca\scb} (y)\ ( e^{-2\p} \d_0 y^\sca\d_0 y^\scb
- e^{2\p}   \d_1 y^\sca  \d_1 y^\scb )  \, ,
\la{bose}
\ee
where $ G_{\sca\scb}$ is the metric of 5-sphere.\foot{Here we
 renamed the (tangent space) indices $\apr,\bpr$
into the coordinate space ones
 $\cal A,B$    for consistency with the notation used
later in  Section 6 ($y^\sca\equiv y^\apr$).}
A peculiarity of the $g_{\mu\nu}$
 gauge choice in \rf{choi}  compared to the usual conformal gauge
is   that  here
the $S^5$ part of the
action is no longer decoupled from the radial $AdS_5$
direction $\phi$.

 (iv)
 The form of the quadratic fermionic  part of the \adss
 superstring action
 expanded near a straight long string  configuration
  along $\p$
 direction of $AdS_5$  was discussed
  in \ci{DGT} using  the  `covariant' $\kappa$-symmetry gauge
  condition $\te^1=\te^2$
 (equivalent result was found  also in the $\te^1= i \gamma_4 \te^2$
 gauge used in \ci{kalram,KT}).
 It is easy to show  that an equivalent fermionic action is found also in
  the present  \lc $\k$-symmetry gauge.
Expanding near the configuration
 $x^0=\tau, \ \p = \s,$ $ x=0,\ y=0$
 (it is easy to check that this   is
 a classical string  solution)
 and choosing the
bosonic  gauge so  that the 2d metric $g_{\mu\nu}$
is equal to the induced ($AdS_2$) metric
 $ ds^2 = { 1 \ov \sigma^2}  ( -d\tau^2 +  d\s^2)$
we find that  the corresponding function $f $  in \rf{solv}
is  then equal to $ \s^{-2}$.
The quadratic   part of the  fermion action \rf{ff3}  becomes
(we redefine the $\eta$ fermions
 by the constant unitary matrix $C'$ in \rf{ff2})
\be
  \int d\tau d\s\    \s^{-2} \big( \te \del_0 \te +
\eta \del_0 \eta -   \eta \del_1 \te \big)   \ .
\ee
Rescaling the fields  \ $\te = \s \te', \ \eta=\s \eta'$
(so that they have $\s$-independent normalization,
$\int d\tau d\s \sg\ \theta \theta = \int d\tau d\s  \theta' \theta'$)
  and integrating by parts  we find
\be
\int d\tau d\s\ \big( \te' \del_0 \te' +
\eta' \del_0 \eta' -   \eta' \del_1 \te'
    -  \s^{-1}  \eta' \te' \big) \ .
\ee
The first three terms here  are  as in the  flat GS action,
while the last term  represents  the $AdS_2$ fermion mass term  which is the same
as found in \ci{DGT}.   Indeed, diagonalizing the action as in \rf{combin}
we get
\be
\int d\tau d\s\ \big( \psi_+ \del_+ \psi_+ + \psi_- \del_- \psi_-
-  \s^{-1}\psi_+ \psi_- \big)
\ , \ee
which is the special case of the  general
form of the quadratic action \rf{ff3}
 with $\del_\s \p$
in the mass term  computed for  $\p= \ln \s$.

%%%%%%%%%%%%%%%%%%%%%%%%%%%%%%%%%%%%%%%%%%
 \subsection{Contents of the rest of the  paper}
 %%%%%%%%%%%%%%%%%%%%%%%%%%%%

 The rest of the paper contains derivation   of the action \rf{oom} and
 related explanations and  technical details.

In Section 2  we start with the case of the  flat space  GS
action   and  illustrate   on this simplest  example
the procedure of light-cone gauge fixing we  shall use in the
$AdS_5\times S^5$ case. We present  a particular
 \lc form  of the GS action
to which our $AdS_5\times S^5$
light-cone gauge fixed  action will reduce in  the flat space limit.

In section 3 we discuss  the basic
superalgebra $psu(2,2|4)$ and
  write down   its   (anti) commutation relations
in the \lc  basis,  corresponding to the   light-cone
 decomposition  (cf. \rf{lcbas})
of the $so(4,2)$ generators.\foot{We shall use the following terminology:
``light cone basis" (or  ``light cone frame")  will refer
to the  decomposition of
 superalgebra  generators,  while  the  ``light cone  gauge"
will refer to the   choice of the
$\kappa$-symmetry gauge.}

In section 4 we  adapt    the original \adss GS action of
 \cite{MT}  to the case of   the
light-cone   basis  of $psu(2,2|4)$.
 The resulting  $\kappa$-symmetric
  action  is written  entirely in terms of Cartan 1-forms
corresponding to
the light-cone  basis and in an arbitrary
(e.g., Wess-Zumino  or
Killing)  parametrization
of the supercoset space.

In section 5  we  fix the
light-cone  $\kappa$-symmetry gauge and  find   the
corresponding Cartan 1-forms.
  These light-cone gauge  1-forms are  given in the
Killing parametrization of the original superspace.

In section 6 we       find the   fermionic \lc gauge
fixed form of  the action  of Section 4.
We present the  action in the
Killing parametrization,   discuss some of its properties, and
also transform  it 
  into the ``4+6"  manifestly 
  $SU(4)$ invariant form  (see \rf{goki},\rf{gowz} and 
  \rf{gokii},\rf{gowzz}).
  We  then  explain  the
transformation of the action into the
Wess-Zumino parametrization  form
which  was  presented  above in \rf{oom},(\ref{ff}),(\ref{qua}).
We  also mention that our results for \adss  case can be easily 
generalized to the $AdS_3 \times S^3$ case.

In Appendix A we discuss the relations between the
$so(4,1)\oplus so(5)$  (or `5+5') basis\foot{We label the basis
by the symmetry algebras under which supercoordinates
are transforming in a linear way.}
 of  the  $psu(2,2|4)$ superalgebra
   used  in \ci{MT}  in the construction of the GS action in \adss
and the    more familiar
  $so(3,1)\oplus su(4)\simeq sl(2,C)\oplus su(4) $  (or `4+6') basis
  (naturally appearing   in the discussion
  of  $\N=4,d=4$ superconformal symmetry of SYM theory).
We  use the later  basis  to identify
 the generators of
the algebra  in    the light-cone (or $so(1,1) \oplus u(1)
 \oplus so(2) \oplus su(4)$)
 basis.
The knowledge of the explicit  relations  between
the generators  in the three  bases
is useful  in order    to find
normalizations   in  the forms of the  string
action  corresponding to  the $so(3,1)\oplus su(4)$
and the light-cone bases.

In Appendix B we explain the transformation of the  \adss
string action from  its original form in the
$so(4,1)\oplus so(5)$ basis  \cite{MT} to the
$so(3,1)\oplus su(4)$ basis and then to the light-cone  basis.
We also discuss some details of derivation of
the light-cone gauge fixed 
action given in  Section 6.

In Appendix C we present  another  version of    the
\adss  superstring
action using  the   ``$S$-gauge" to fix the
$\kappa$-symmetry ($S$ refers to
the  conformal supersymmetry generator).  In this gauge
all of the superconformal $\eta$-fermions are gauged away.

%%%%%%%%%%%%%%%%%%%%%%%%%%%%%%%%%%%%%%%%%%%%%%%%%%%%%%
%%%%%%%%%%%%%%%%%%%%%%%%%%%%%%%%%%%%%%%%%%%
\newsection{ Light cone $\kappa$-symmetry   gauge fixing in flat space}
%%%%%%%%%%%%%%%%%%%%%%%%%%%%%%

It is useful first to discuss the case of \lc gauge  fixing
in the standard flat space GS action. This allows
to explain   in the simplest  setting
 the procedure
of light-cone gauge fixing we are going to  follow   in   the case of
$AdS_5\times S^5$. In particular, we shall discuss
 the split of supercoordinates which is
closely related to the one
we will use in the $AdS_5\times  S^5$  case, and obtain
 the form of the GS action to which our $AdS_5\times  S^5$
 \lc action will
reduce in the flat space limit.

We  start with the flat GS action \ci{GS}
in the form \ci{HM}
\be
I_0=-\frac{1}{2}\int_{\del {M_3}}  d^2\sigma\  \sqrt{g}\ g^{\vm\vn}\
 L_{\vm}^{\hat A}  L_{\vn}^{\hat A}
+  {\rm i}\int_{M_3}
 s^{IJ}   L^{\hat A}\wedge  (\bar{L}^I\Gamma^{\hat A}
\wedge  L^J)
\ ,
\la{actif}
\ee
where   $s^{IJ}\equiv {\rm diag}( 1,-1)$  ($I,J=1,2$)
and
 ${ 2\pi \alpha'}=1$.  The 2-d metric
$g_{\vm\vn}$ ($\vm,\vn=0,1$)  has  signature
$(-+)$, and $g\equiv - \det g_{\vm\vn}$.

The left-invariant Cartan  1-forms are defined on the
type IIB coset superspace
defined as
[10-d super Poincare group]/[$SO(9,1)$ Lorentz group]
%$$L^{\hat{A}} = dX^M L^{\hat{A}}_{M}\ ,  \qquad \ \ \
%X^M=(x, \theta)$$
\begin{equation}\label{expan1}
G^{-1}dG =
L^{\hat{A}}P_{\hat{A}} + L^I Q_I\ , \ \ \ \ \
  L^{\hat{A}} = dX^M L^{\hat{A}}_{M}\ ,  \qquad \
X^M=(x, \theta) \ ,
\end{equation}
where $G= G({x,\theta})$ is an appropriate coset representative.
A specific choice of $G(x,\theta)$ commonly used is
\be
G({x,\t}) = {\rm exp} ( x^{\hat A} P_{\hat A}  + \t^I Q_I)\,,
\qquad [P_{\hat A}, P_{\hat B}]=0\,,
\ \ \  \{ Q_I, Q_J\}
=-2{\rm i}  \delta_{IJ} (\CC\Gamma^{\hat A}) P_{\hat A}\,,
\la{supf}
\ee
and  thus  the coset space vielbeins  defined by (\ref{expan1}) are given
by
\be
L^{\hat A} = d x^{\hat A}  - i \bar \t^I \G^{\hat A}d\t^I \,, 
\qquad
L^I = d \t^I \ .
\la{flat}
\ee
$\theta^I$  are two left Majorana-Weyl 10-d spinors.
The  explicit  2-d form of the GS action \ci{GS}
$$
I_0=\int d^2 \sigma\ {\cal L}_0 =
  \int d^2 \sigma\  \bigg[ - \ha
\sqrt{g} g^{\vm\vn} (\del_\vm x^{\hat A} -
{\rm i} \bar \t^I \G^{\hat A} \del_\vm \t^I)
(\del_\vn x^{\hat A} -
{\rm i} \bar \t^J \G^{\hat A} \del_\vn \t^J)
$$ \be
-{\rm i} \ep^{\vm\vn} s^{IJ}   \bar \t^I \G^{\hat A} \del_\vn  \t^J
(\del_\mu  x^{\hat A} -  \ha {\rm i}   \bar \t^K \G^{\hat A}  \del_\vm \t^K)
 \bigg] \ .
\la{GRE}
\ee
One usually  imposes the $\kappa$-symmetry
\lc gauge by starting with  the component  form of action given by
(\ref{GRE}). It turns out to be    cumbersome
to  generalize this procedure to the case of
 strings in $AdS_5\times  S^5$.
 It is more  convenient  to first impose the  \lc    gauge
at the level of the Cartan forms $L^{\hat{A}}$, $L^I$ and
then use  them
 in   the action taken  in its   general  form (\ref{actif}).
The  \lc gauge form of  $L^{\hat A}$ is
\begin{equation}\label{trcarfor}
L^+=dx^+\,,
\qquad
L^- = dx^- - {\rm i}\bar{\theta}^I\Gamma^- d\theta^I\,,
\qquad
L^N=dx^N\,,
\qquad
N=1,\ldots,8 \ ,
\end{equation}
where $\theta^I$ are subject to the
\lc gauge condition $\Gamma^+\theta^I=0$.\foot{ The
transverse bosonic Cartan forms $L^N$ in (\ref{trcarfor}) should not be
confused with fermionic ones $L^I$.}
Inserting these  expressions  into
action (\ref{actif}) we get
\begin{equation}
{\cal L}={\cal L}_{kin}+{\cal L}_{WZ}  \ ,
\la{aca}
\end{equation}
\begin{equation}
{\cal L}_{kin}
=\sqrt{g}g^{\vm\vn}(
-\partial_\vm x^+\partial_\vn x^- -
\frac{1}{2}\partial_\vm x^N\partial_\vn x^N
+{\rm i}\partial_\vm x^+\bar{\theta}^I\Gamma^-\partial_\vn \theta^I) \ ,
\end{equation}
\begin{equation}
{\cal L}_{WZ}= -{\rm i}\epsilon^{\vm\vn}s^{IJ}\partial_\vm x^+
\bar{\theta}^I\Gamma^-\partial_\vn \theta^J    \ .
\end{equation}
Next,
let us do the   ``$5+5$" splitting of
the 10-d Clifford algebra generators, the charge conjugation matrix  ${\cal C}$
 and  the supercoordinates
\be
\G^\aA=\gamma^\aA\times I\times \sigma_1 \,,
\quad
\G^\apr=I\times\gamma^\apr\times \sigma_2\,,
\quad  \CC =C\times C^\prime \times {\rm i}\sigma_2\,,
\quad
\theta^I=
\left(
\begin{array}{c}
\theta^{I\alpha i}
\\
0
\end{array}
\right)\ ,
\ee
where $I$ is  $4\times 4$ unit matrix,  $\sigma_n$ are Pauli
matrices,  $\alpha=1,2,3,4$ and $ \ i=1,2,3,4$.
Let us also   introduce the complex coordinates
\be
\theta^q\equiv\frac{1}{\sqrt{2}}(\theta^1+{\rm i}\theta^2) \ ,
\ee
and use the parametrization
\be
\theta^{q\alpha i}
= \frac{v}{2} \left(
\begin{array}{c}
\eta^{-i}
\\
\eta^{+i}
\\
-{\rm i}\theta^{+i}
\\
{\rm i} \theta^{-i}
\end{array}\right)     \ , \ \ \ \ \ \ \ \      v\equiv 2^{1/4}\ .
\la{hyt}
\ee
Decompositions of $so(4,1)$ $\gamma$-matrices we use may be found in
Appendix A (see (\ref{gamdec})).
The \lc  gauge
\be
\Gamma^+\theta^I=0\ , \ \ \ \ \ \ \
\Gamma^\pm  \equiv { 1 \over \sqrt{2}}
(\Gamma^3 \pm \Gamma^0)      \ ,
\ee
 leads to
\be
\theta^{+i}=\eta^{+i}=0\ .  \la{oldi}
\ee
  Changing sign $x^{\hat{A}}\rightarrow - x^{\hat{A}}$, using the notation
  \be
  \theta^i\equiv \theta^{-i}\ , \ \ \ \ \ \ \ \
\eta^i\equiv \eta^{-i}   \ ,  \ \ \ \ \ \ \ \ \ \
\theta_i = (\theta^i)^\dagger\ , \ \ \
\eta_i = (\eta^i)^\dagger  \ ,
\ee
 and inserting the above  decomposition  into the
  action \rf{aca} we  finally get
  the following expressions for the kinetic and WZ parts of the
  \lc gauge  flat space GS  Lagrangian
\begin{eqnarray}\label{gskin2}
{\cal L}_{kin}
=\sqrt{g}g^{\vm\vn}
\Bigl[
-\partial_\vm x^+ \partial_\vn x ^-
-\frac{1}{2}\partial_\vm x^N\partial_\vn x^N
-\partial_\vm x^+(
\frac{{\rm i}}{2}\theta_i\partial_\vn \theta^i
+\frac{{\rm i}}{2}\eta_i\partial_\vn\eta^i+h.c.)\Bigr]  \ ,
\end{eqnarray}
\begin{equation}\label{gswz2}
{\cal L}_{WZ}
=\epsilon^{\vm\vn}
\partial_\vm x^+ \eta^i C_{ij}^\prime \partial_\vn\theta^j+h.c.
\end{equation}
It is to this form of the flat  GS action
 that  our \lc  \adss action
will reduce in  the
flat space limit.
A  characteristic feature of this
parametrization is
 that while the kinetic term is diagonal
in $\theta$'s and $\eta$'s, they are mixed in the WZ term.

%%%%%%%%%%%%%%%%%%%%%%%%%%%%%%%%%%%%%%%%%%%%%%%%%%%%%%%%%%%%
\newsection{$psu(2,2|4)$ superalgebra in the light cone
basis}
%%%%%%%%%%%%%%%%%%%%%%%%%%%%%%%%%%%%%%%%%%%%%%%%%%%%

The  superalgebra $psu(2,2|4)$  which is the algebra of isometry
transformations of  $AdS_5\times S^5$ superspace
plays the
central  role in the construction of the GS action in \adss \ci{MT}.
In this section we shall present the form of this  algebra
which will be used in the present paper.
  The even part of $psu(2,2|4)$
   is the sum of  the algebra $so(4,2)$ which is the
isometry algebra of $AdS_5$ and the algebra $so(6)$ which is the  isometry
algebra of $S^5$.
 The odd part  consists of 32 supercharges
corresponding to  32 Killing spinors in $AdS_5\times S^5$
vacuum \ci{john}
of type IIB supergravity
(see \cite{GM1,FI,HLS};  for recent developments  in
 representation theory see \cite{twow}).

We  shall  use the form of the  basis of
$so(4,2)$ sub-algebra
 implied by its interpretation
 as conformal algebra in 4 dimensions. The generators are then
 called   translations
$P^a$, conformal boosts $K^a$, dilatation $D$ and Lorentz rotations
$J^{ab}$    and  satisfy the  standard commutation relations
\begin{equation}
\label{comrel1}
[P^a,J^{bc}]=\eta^{ab}P^c-\eta^{ac}P^b,
\quad
[K^a,J^{bc}]=\eta^{ab}K^c-\eta^{ac}K^b,
\quad
[P^a,K^b]=\eta^{ab}D-J^{ab},
\end{equation}
\begin{equation}\label{comrel2}
[D,P^a]=-P^a,
\quad
[D,K^a]=K^a,
\qquad
[J^{ab},J^{cd}]=\eta^{bc}J^{ad}+ 3\hbox{ terms}\,,
\end{equation}
where $\eta^{ab}=(-,+,+,+)$  and
$a,b,c,d=0,1,2,3$.
 In the light cone   basis  \rf{lcbas}
  we have the following generators:
  \be
 J^{+-}\ , \ \ \  J^{\pm x}\ , \ \ \
  J^{\pm\bar x} \ , \ \ \   
J^{x\bar x} \ , \ \ \ P^\pm   \ , \ \ \    P^x   \ , \ \ \
P^{\bar x}  \ , \ \ \    K^\pm \ , \ \ \   K^x   \ , \ \ \
 K^{\bar x}
\ .
 \la{gene}
\ee
To simplify the  notation  we shall set
\be P\equiv P^x\ , \ \ \ \
\bar{P} = P^{\bar x}\ , \ \ \ \ K\equiv K^x\ ,
\ \ \ \ \  \bar{K}=K^{\bar x}\ .
\ee
The light cone form of $so(4,2)$ algebra commutation relations can be
obtained from (\ref{comrel1})
using that the light cone metric has the
following  elements $\eta^{+-}=\eta^{-+}=1$,
$\eta^{x\bar{x}}=\eta^{\bar{x}x}=1$.

 In this paper  the  $so(6)$
algebra  will be  interpreted as   $su(4)$ one
($i,j,k,l=1,2,3,4$)
\begin{equation}\label{su4com}
[J^i{}_j,J^k{}_n]=\delta^k_jJ^i{}_n
-\delta^i_n J^k{}_j\,.
\end{equation}

To describe the odd part of $psu(2,2|4)$ superalgebra we introduce
32 supercharges $Q^{\pm i}$, $Q_i^\pm$, $S^{\pm i}$, $S_i^\pm$.
They carry
 the  $D$, $J^{+-}$ and $J^{x\bar x}$ charges, as follows from the
 structure of the algebra. The
 commutation relations of the
 supercharges with the dilatation $D$
\begin{equation}
[D,Q^{\pm i}]=-\frac{1}{2}Q^{\pm i}\,,
\quad
[D,Q^\pm_i]=-\frac{1}{2}Q^\pm_i\,,
\qquad
[D,S^{\pm i}]=\frac{1}{2}S^{\pm i}\,,
\quad
[D,S^\pm_i]=\frac{1}{2}S^\pm_i\,,
\end{equation}
allow to interpret   $Q$'s as the standard  supercharges
of the super Poincar\'e subalgebra and $S$'s  as 
the  conformal supercharges.
The supercharges with superscript $+$ ($-$) have positive (negative)
$J^{+-}$ charge
$$
[J^{+-},Q^{\pm i}]
=\pm\frac{1}{2}Q^{\pm i}\,,
\quad  \ \
[J^{+-},Q^\pm_i]
=\pm\frac{1}{2}Q^\pm_i\,,
$$
$$
[J^{+-},S^{\pm i}]
=\pm\frac{1}{2}S^{\pm i}\,,
\quad   \ \
[J^{+-},S^\pm_i]
=\pm\frac{1}{2}S^\pm_i\,.
$$
The $J^{x\bar x}$ charges are fixed by the commutation relations
\begin{equation}
[J^{x\bar{x}},Q^{\pm i}]=\pm\frac{1}{2}Q^{\pm i}\,,
\quad
[J^{x\bar{x}},Q^\pm_i]=\mp\frac{1}{2}Q^\pm_i\,,
\end{equation}
\begin{equation}
[J^{x\bar{x}},S^\pm_i]=\pm\frac{1}{2}S^\pm_i\,,
\quad
[J^{x\bar{x}},S^{\pm i}]=\mp\frac{1}{2}S^{\pm i}\,.
\end{equation}
The transformation properties of the $Q$ supercharges with respect to $su(4)$
subalgebra are  determined by
$$
[Q^\pm_i,J^j{}_k]=\delta_i^jQ^\pm_k-\frac{1}{4}\delta^j_kQ^\pm_i\,,
\qquad    \ \
[Q^{\pm i},J^j{}_k]=-\delta^i_k Q^{\pm j}
+\frac{1}{4}\delta^j_kQ^{\pm i}   \ ,
$$
 with the same relations for the $S$ supercharges.
Anticommutation relations between the supercharges are
\begin{equation}
\{Q^{\pm i},Q_j^\pm\}=\mp{\rm i} P^\pm\delta^i_j\,,
\quad     \ \ \
\{Q^{+i},Q_j^-\}=-{\rm i}P\delta^i_j\,,
\end{equation}
\begin{equation}
\{S^{\pm i},S_j^\pm\}=\pm {\rm i}K^\pm\delta^i_j\,,
\quad \ \
\{S^{-i},S_j^+\}={\rm i}K\delta^i_j\,,
\end{equation}
$$
\{Q^{+i},S^+_j\}=-J^{+x}\delta^i_j\,,
\quad    \ \
\{Q^{-i},S^-_j\}=-J^{-\bar{x}}\delta^i_j\,,
$$
$$
\{Q^{\pm i},S^\mp_j\}=\frac{1}{2}(J^{+-}+J^{x\bar{x}}\mp D)\delta_j^i
\ \mp \ J^i{}_j\,.
$$
The remaining commutation relations between odd  and even  generators
 have  the following form
$$
[Q^{-i},J^{+x}]=-Q^{+i}\,,
\quad
[S^{-i},J^{+\bar{x}}]=-S^{+i}\,,
\quad
[Q^{+i},J^{-\bar{x}}]=Q^{-i}\,,
\quad
[S^{+i},J^{-x}]=S^{-i}\,,
$$
$$
[S^\mp_i,P^\pm]={\rm i}Q^\pm_i\,,
\quad  \ \
[S^-_i,P]={\rm i}Q^-_i\,,
\quad     \ \
[S^+_i,\bar{P}]=-{\rm i}Q^+_i\,,
$$
$$
[Q^{\mp i},K^\pm]=-{\rm i}S^{\pm i}\,,
\quad   \ \
[Q^{-i},K]=-{\rm i}S^{-i}\,,
\quad   \ \
[Q^{+i},\bar{K}]={\rm i}S^{+i}\,.
$$
The  generators are subject to  the following hermitean
conjugation  conditions
$$
(P^{\pm})^\dagger=-P^\pm,
\quad
P^\dagger=-\bar{P},
\quad
(K^\pm)^\dagger=-K^\pm,
\quad
K^\dagger=-\bar{K},
$$
$$
(J^{\pm x})^\dagger=-J^{\pm \bar{x}}\,,
\quad
(J^{+-})^\dagger=-J^{+-}\,,
\quad
(J^{x\bar{x}})^\dagger=J^{x\bar{x}}\,,
\quad
D^\dagger=-D\,,
\quad
(J^i{}_j)^\dagger=J^j{}_i\,.
$$
\begin{equation}\label{herconrul3}
(Q^{\pm i})^\dagger=Q^{\pm}_i,  \ \
\quad           \ \
(S^{\pm i})^\dagger=S^{\pm}_i\,,
\end{equation}
All the remaining nontrivial (anti)commutation relations of $psu(2,2|4)$
superalgebra  may be obtained by using these hermitean conjugation rules
and the (anti)commutation relations given above.

%%%%%%%%%%%%%%%%%%%%%%%%%%%%%%%%%%%%%%%%%%%%%
\newsection{ Light cone basis form of \adss   string action }
%%%%%%%%%%%%%%%%%%%%%%%%%%%%%%%%%%%%%%%%%%%%%%%%%%%

Superstring action in $AdS_5\times S^5$   \cite{MT}
has the same structure as  the flat space  GS  action \rf{actif}
\begin{equation}\label{action}
I=\int_{\partial M_3} {\cal L}_{kin}
+\int_{M_3} {\rm i}{\cal H}    \ .
\end{equation}
In \cite{MT} the  Cartan forms in terms of which the action is written
were   given  in  the
$so(4,1)\oplus so(5)$ basis of $psu(2,2|4)$.
 This is the  basis that allows
to present the \adss GS action   in  the  form similar to the 
one in the 
flat space. Our present  goal   is to rewrite the action in
the light-cone basis discussed in the previous section
and then (in the next Section)  to impose  a $\kappa$-symmetry
light-cone gauge.
We shall use the   conformal algebra  and
light-cone frame notation.

The Cartan 1-forms in the light-cone basis are defined by
\begin{eqnarray}\label{carfor}
G^{-1}dG
&=&
L_\ssmP^aP^a+L_\ssmK^aK^a+L_\ssmD D  
+\frac{1}{2}L^{ab}J^{ab} + L^i{}_j J^j{}_i
\nonumber\\
&& +\  L_\ssmQ^{-i}Q_i^+ + L_{\ssmQ\, i}^-Q^{+i}
+ L_\ssmQ^{+i}Q_i^- + L_{\ssmQ\,i}^+ Q^{-i}
\nonumber\\
&& +\ L_\ssmS^{-i}S_i^+ + L_{\ssmS\, i}^- S^{+i}
+ L_\ssmS^{+i} S_i^- + L_{\ssmS\,i}^+ S^{-i}  \ ,
\end{eqnarray}
where  $G$ is a coset representative in $PSU(2,2|4)$.
Let us define also the following combinations
\begin{equation}\label{hll}
\hat{L}^a\equiv L_\ssmP^a-\frac{1}{2}L_\ssmK^a\,,
\qquad
L^\apr\equiv -\frac{\rm i}{2}(\gamma^\apr)^i{}_j L^j{}_i\,,
\qquad
(\Csp L)_{ij}\equiv \Csp_{ik} L^k{}_j   \ .
\end{equation}
Then   the kinetic term in \rf{action}
takes the form
\begin{equation}\label{lkin1}
{\cal L}_{kin}
=-\frac{1}{2}\sqrt{g}g^{\vm\vn}
\Bigl(\hat{L}_\vm^a\hat{L}_\vn^a
+L_{\ssmD\vm}L_{\ssmD\vn}
+L_\vm^\apr L_\vn^\apr\Bigr) \ ,
\end{equation}
while the 3-form ${\cal H}$ in the WZ term  can be written as
(we suppress  the signs of exterior products of 1-forms)
\begin{equation}
{\cal H}={\cal H}_{AdS_5}^q+{\cal H}_{S^5}^q - h.c. \ ,
\end{equation}
\begin{eqnarray}
{\cal H}_{AdS_5}^q 
&=&-\frac{{\rm i}}{\sqrt{2}}(\hat{L}^+L_\ssmS^{-i}\Csp_{ij}L_\ssmQ^{-j}
+\hat{L}^-L_\ssmQ^{+i}\Csp_{ij} L_\ssmS^{+j}
+\hat{L}^xL_\ssmS^{-i}\Csp_{ij} L_\ssmQ^{+j}
+\hat{L}^{\bar{x}} L_\ssmS^{+i}\Csp_{ij}L_\ssmQ^{-j})
\nonumber\\
&+&
\frac{1}{\sqrt{2}}L_\ssmD
(\frac{1}{2}L_\ssmS^{+i}\Csp_{ij} L_\ssmS^{-j} +L_\ssmQ^{-i}\Csp_{ij}
L_\ssmQ^{+j})  \ ,
 \label{wz1}
\\
{\cal H}_{S^5}^q
&=&
\!\!
\frac{1}{2\sqrt{2}}[L_\ssmS^{+i}(\Csp L)_{ij}L_\ssmS^{-j}
-L_\ssmS^{-i}(\Csp L)_{ij}L_\ssmS^{+j}]
+\frac{1}{\sqrt{2}}[L_\ssmQ^{+i}(\Csp L)_{ij}L_\ssmQ^{-j}
-L_\ssmQ^{-i}(\Csp L)_{ij}L_\ssmQ^{+j}]
\nonumber
\end{eqnarray}
 Derivation of these  expressions from
 the original ones  given in
\cite{MT} may be found in  Appendix B.

%%%%%%%%%%%%%%%%%%%%%%%%%%%%%%%%%%%%%%%%%%%%%%%%%
\newsection{Coordinate parametrization  of Cartan forms  and fixing
 the light-cone $\kappa$-symmetry gauge}
 %%%%%%%%%%%%%%%%%%%%%%%%%%%%%%%%%%%%%%%%%%%%%%%%

 To represent the Cartan 1-forms in terms of the
 even and odd coordinate      fields  we shall
 start with the following supercoset representative    (cf. \rf{supf})
\be
G= \ g_{x,\theta} \  g_\eta \ g_y\ g_\phi \ ,
\la{kilgau}
\ee
 where
 \be
 g_{x,\theta} =
\exp(x^aP^a +\theta^{-i} Q_i^+ +\theta_i^-Q^{+i}+\theta^{+i} Q_i^-
+\theta_i^+Q^{-i}) \ , \ee
 \be
 g_\eta =
\exp(\eta^{-i}S_i^+ + \eta_i^- S^{+i}+\eta^{+i}S_i^- +
\eta_i^+S^{-i})\ ,
\ee
and
$g_\phi$ and $g_y$ depend on the  radial $AdS_5$ coordinate $\phi$
and $S^5$ coordinates $y^\apr$ respectively:
\begin{eqnarray}
 &&
g_\phi \equiv \exp(\phi D)\ , \\
&&
g_y\equiv \exp(y^i{}_jJ^j{}_i)  \ ,\ \ \ \  \ \ \ \ \ \
 y^i{}_j\equiv \frac{{\rm i}}{2}(\gamma^\apr)^i{}_j y^\apr \ .
\end{eqnarray}
Choosing the parametrization of the
 coset representative in the  form (\ref{kilgau})
 corresponds to what is   usually    referred
to as   ``Killing gauge"   in superspace.

Since the
supercharges transform in the fundamental representation
of $su(4)$ the
corresponding  fermionic coordinates $\theta$'s  and $\eta$'s
 also transform
in the fundamental representation of $su(4)$.

  The above
expressions provide the  definition of the Cartan forms
 in the light-cone basis.
 Let us   further specify  these expressions  by setting
 to zero some of the fermionic coordinates
  which corresponds to fixing
   a  particular  $\kappa$-symmetry gauge.
Namely, we shall
 fix the $\kappa$-symmetry by putting to
 zero all the  Grassmann coordinates which carry
positive $J^{+-}$ charge (cf. \rf{oldi}):
\begin{equation}\label{kapfixt}
\theta^{+i}=\theta_i^+=\eta^{+i}=\eta_i^+=0 \  .
\end{equation}
To simplify the notation  we shall set in what follows
\be
\theta^i\equiv \theta^{-i}\,,
\qquad
\theta_i\equiv \theta_i^-\,,
\qquad
\eta^i\equiv \eta^{-i}\,,
\qquad
\eta_i\equiv \eta_i^-\, .
\ee
As a result,
the  $\kappa$-symmetry fixed  form of the
coset representative \rf{kilgau}  is
\be\label{kapfixG}
  G_{g.f.} =\   (g_{x,\theta})_{g.f.} \  (g_\eta)_{g.f.}\  g_y\  g_\phi \ ,
\ee
\begin{eqnarray}
&&
 (g_{x,\theta})_{g.f.}= \exp(x^aP^a +\theta^i Q_i^+ +\theta_iQ^{+i})  \ ,
\\
&&
 (g_\eta)_{g.f.}=\exp(\eta^iS_i^+ + \eta_iS^{+i})  \ .
\end{eqnarray}
Plugging this $G_{g.f.}$  into  (\ref{carfor}) we get
the $\kappa$-symmetry gauge fixed expressions for the Cartan
1-forms
\begin{eqnarray}
\label{carfor1}
&&
L_\ssmP^+ = e^\phi dx^+   \ , \ \ \ \   \
  L_\ssmP^-=e^\phi (dx^-
- \frac{{\rm i}}{2}\tilde{\theta}^i\tilde{d\theta}_i
-\frac{{\rm i}}{2}\tilde{\theta}_i \tilde{d{\theta^i}})   \ ,
\\
&&
L_\ssmP^x =e^\phi dx  \ , \ \ \ \ \ \ \ \
 L_\ssmP^{\bar{x}} =e^\phi d\bar{x}  \ ,
\\
&&
L_\ssmK^-=e^{-\phi}\bigl[\frac{1}{4}(\tilde{\eta}^2)^2 dx^+
+\frac{{\rm i}}{2}\tilde{\eta}^i \tilde{d\eta}_i
+\frac{{\rm i}}{2}\tilde{\eta}_i \tilde{d{\eta^i}}\bigr]   \ ,
\\
&&
L_\ssmD=d\phi \ ,  \\
&&
L^i{}_j= (dUU^{-1})^i{}_j+{\rm i}(\tilde{\eta}^i\tilde{\eta}_j
-\frac{1}{4}\tilde{\eta}^2\delta_j^i)dx^+   \ ,  \la{carforij}\\
 &&
L_\ssmQ^{-i}=e^{\phi/2}(\tilde{d\theta}^i +{\rm i}\tilde{\eta}^i dx) \ ,
\ \ \ \ \ \  \
L_{\ssmQ i}^-=e^{\phi/2}(\tilde{d\theta}_i - {\rm i}\tilde{\eta}_i
d\bar{x}) \ ,
\\
&&
L_\ssmQ^{+i}=-{\rm i}e^{\phi/2}\tilde{\eta}^i dx^+
\ , \ \ \ \ \ \ \
L_{\ssmQ i}^+={\rm i}e^{\phi/2}\tilde{\eta}_i dx^+ \ ,
\\
&&
L_\ssmS^{-i}=e^{-\phi/2}(\tilde{d\eta}^i
+\frac{{\rm i}}{2}\tilde{\eta}^2\tilde{\eta}^i dx^+)
\ , \ \ \ \ \ \ \
L_{\ssmS i}^- = e^{-\phi/2}(\tilde{d\eta}_i
-\frac{{\rm i}}{2}\tilde{\eta}^2\tilde{\eta}_i dx^+)
\ , \label{carfor2}
\end{eqnarray}
where $\td \eta^2 \equiv \td\eta^i \td\eta_i$. 
All the remaining forms are equal to zero.
We  have  introduced the notation
\begin{equation}\label{tilthe}
\tilde{\theta}^i \equiv  U^i{}_j \theta^j\,,
\qquad
\tilde{\theta}_i \equiv  \theta_j (U^{-1})^j{}_i  \ ,
\end{equation}
\begin{equation}\label{tildthe}
\tilde{d{\theta^i}}\equiv U^i{}_j d\theta^j\,,
\qquad
\tilde{d\theta}_i \equiv  d\theta_j (U^{-1})^j{}_i      \  ,
\end{equation}
and similar ones  for   for $\eta$.
 Note that
$\tilde{\theta}^2 =\theta^2$ and
$\tilde{\theta}\tilde{d\theta}=\theta d\theta$.
The matrix $U\in SU(4) $ is defined  in terms of
the $S^5$ coordinates  $y^i_{\ j}$  or $y^\apr$
by (\ref{uuu}),(\ref{dere}). It can be   written  explicitly as
\be
U=\cos{|y|\over 2 } +{\rm i}\gamma^\apr n^\apr \sin{|y|\over 2 }
\,,
\qquad
|y|\equiv\sqrt{y^\apr y^\apr}\,,
\quad
n^\apr \equiv  \frac{y^\apr}{|y|}  \ .   \la{uuuu}
\ee

%%%%%%%%%%%%%%%%%%%%%%%%%%%%%%%%%%%%%%%%%%%%%%%%%%%%%%%%%%%
\newsection{\adss string  action in the  light-cone gauge}
%%%%%%%%%%%%%%%%%%%%%%%%%%%%%%%%%%%%%%%%%%%%%%%%%%%

Plugging the above expressions \rf{carfor1}--\rf{carfor2}
for  into the  action (\ref{action}) we
get the following result  for the  light-cone gauge fixed
superstring Lagrangian  in terms of the light
cone supercoset coordinates

\begin{eqnarray}\label{actkin2}
{\cal L}_{kin}
=  &&
\sqrt{g}g^{\vm\vn}\left(
-e^{2\phi}(\partial_\vm x^+ \partial_\vn x ^-
+ \partial_\vm x\partial_\vn\bar{x})
-\frac{1}{2}\partial_\vm \phi\partial_\vn \phi
-\frac{1}{2}e_\vm^\apr e_\vn^\apr \right.
\nonumber\\
&+&
%\sqrt{g}g^{\vm\vn}
\partial_\vm x^+
\Bigl[ \frac{{\rm i}}{2}  e^{2\phi}(\theta^i\partial_\vn \theta_i
+\theta_i\partial_\vn \theta^i)
+\frac{{\rm i}}{4}(\eta^i\partial_\vn \eta_i
+\eta_i\partial_\vn \eta^i) 
+\frac{1}{2}\tilde{\eta}_ie_\vn^i{}_j\tilde{\eta}^j \Bigr]
\nonumber\\
&+&
%\sqrt{g}g^{\vm\vn}
\left. \frac{1}{8}\ \partial_\vm x^+\partial_\vn x^+
\bigl[(\eta^2)^2
-(\tilde{\eta}_i(\gamma^\apr)^i{}_j\tilde{\eta}^j)^2 \bigl] \right) \ ,
\end{eqnarray}
\begin{equation}\label{actwz2}
{\cal L}_{WZ}
=-\frac{\epsilon^{\vm\vn}}{\sqrt{2}}
e^\phi\partial_\vm x^+ \tilde{\eta}^i C_{ij}^\prime
(\tilde{\partial_\vn\theta}^j+{\rm i}\tilde{\eta}^j \partial_\vn x)+h.c.
\end{equation}
The kinetic terms are obtained
 in a  straightforward way. Details of
derivation of the WZ part  are given  in Appendix B.

Few remarks are in order.

(i) In the flat space limit this  action
(after an appropriate  rescaling of fermionic variables 
given in (\ref{rescal}))
reduces to the  GS  light-cone $\kappa$-symmetry gauge fixed action
represented in the form
 (\ref{gskin2}),(\ref{gswz2}). In the  particle  theory
 limit   $\alpha'\to 0$  (corresponding to  keeping only the
  $\tau$ dependence of the fields
 and omitting the WZ term)
 this action reduces (after an
 appropriate  bosonic light-cone  gauge fixing
 and rescaling of some fermionic variables)
 to the light-cone  action of  a superparticle
 propagating  in $AdS_5\times S^5$
\cite{metsii}.\foot{Ref. \cite{metsii}  found the
 Hamiltonian for the superparticle in \adss (see eq. (12) there).
 The action  is obtained  from the Hamiltonian in the usual way.}

(ii) The kinetic terms for the
 fermionic coordinates  have  manifest linear $su(4)$
invariance. In the remaining terms this symmetry is not manifest
and is not realized linearly.

(iii) Since the WZ term depends on $\theta$ through its derivative,
it   is invariant under a shift of $\theta$. To
maintain this  invariance in the kinetic terms
 the shifting of $\theta$ should
be supplemented, as usual, by an
appropriate transformation of $x^-$.
The action is invariant under shifts of the bosonic coordinates  $x^a$
along the boundary directions.

(iv) As in the superparticle case \cite{metsii,metsiii}
this  action is
quadratic in half of the fermionic coordinates ($\theta$)
 but of higher order (quartic)  in the  another half
 ($\eta$).
 It  was   the desire to split the fermionic variables in
 such $\theta$'s and
$\eta$'s that motivated  our choice
of the  supercoset parametrization  in
(\ref{kapfixG}).

(v) The action contains  the  terms like $(\eta^2)^2$ and
$\eta_i e^i{}_j \eta^j$ which in the superparticle case played important role
\cite{metsii} in  establishing
 the AdS/CFT correspondence. These
terms should  also play a similar important role
in  formulating the  AdS/CFT correspondence at the string theory level.

The fermionic variables $\theta$ and $\eta$ as
 defined above in \rf{kilgau}
 have opposite
conformal dimensions.
 It is convenient,  however,  to use the variables with the
same conformal dimensions.\footnote{The light-cone formulation of the
superparticle in $AdS_5\times S^5$ \cite{metsii,metsiii} used similar
 Grassmann variables with the same conformal dimensions}
 To achieve this
we rescale $\eta$ as follows

\be\label{rescal}
\eta^i \rightarrow \sqrt{2}e^\phi \eta^i\,,
\qquad
\eta_i \rightarrow \sqrt{2}e^\phi \eta_i  \ .
\ee
To get convenient sign in front of kinetic terms of fermions 
we change sign $x^a\rightarrow -x^a$. 
Then the Lagrangian (\ref{actkin2}),(\ref{actwz2})
may be written as follows

$$
{\cal L}_{kin}
=
\sqrt{g}g^{\vm\vn}\Bigl[
-e^{2\phi}(\partial_\vm x^+ \partial_\vn x ^-
+ \partial_\vm x\partial_\vn\bar{x})
-\frac{1}{2}\partial_\vm \phi\partial_\vn \phi
-\frac{1}{2}G_{\sca\scb}(y) D_\vm y^\sca D_\vn y^\scb\Bigr]
$$
\be
- \ \frac{{\rm i}}{2} \sqrt{g}g^{\vm\vn}
e^{2\phi}\partial_\vm x^+
\Bigl[\theta^i\partial_\vn \theta_i
+\theta_i\partial_\vn \theta^i
+\eta^i\partial_\vn \eta_i
+\eta_i\partial_\vn \eta^i 
 +  {\rm i}  e^{2\phi}\partial_\vn x^+(\eta^2)^2\Bigr]\ ,
\label{actkin3}
\ee
\begin{equation}\label{actwz3}
{\cal L}_{WZ}
=\epsilon^{\vm\vn}
e^{2\phi}\partial_\vm x^+ \eta^i C_{ij}^U
(\partial_\vn\theta^j - {\rm i}\sqrt{2}e^\phi\eta^j
\partial_\vn x)+h.c.
\end{equation}
Here $G_{\sca\scb}$ is the  metric of  5-sphere.\foot{We introduced the
coordinate $S^5$ indices ${\cal A,B}=1,...,5$
(to be  distinguished  from the tangent space indices
$\apr,\bpr$)
and set $y^{\cal A}= \delta^{\cal A}_{\apr} y^{\apr}$.}
The matrix $C_{ij}^U$ and the differential
$D_\mu y^\sca$ are defined by
\begin{equation}\label{dya}
D_\mu  y^\sca = \del_\mu  y^\sca - 2{\rm i}\eta_i
(V^\sca)^i{}_j\eta^j e^{2\phi} \del_\mu  x^+  \ ,
\end{equation}
\be\label{CU}
C_{ij}^U\equiv
U^l{}_i C_{lk}^\prime U^k{}_j\,,
\qquad
C_{ij}^U=C_{ij}^\prime \cos|y| 
+{\rm i}(C^\prime\gamma^\apr)_{ij}n^\apr\sin |y|
\ee
where  $(V^\sca)^i{}_j$ are  the components of
the Killing vectors
$(V^\sca)^i{}_j  \partial_{y^\sca}$ of $S^5$ \ 
 ($\partial_{y^\sca}  = {\del\over \del y^\sca}$).

Note that $x^+$ enters the action only  through  the  combination
$ e^{2\phi}\partial_\vm x^+$.
An  attractive feature of  this  representation is that the
terms in \rf{actkin2}  involving
 $\tilde{\eta}_i(\gamma^\apr)^i{}_j\tilde{\eta}^j$
 are now collected in  the second term in the derivative
(\ref{dya}) and  thus have  a
natural  geometrical  interpretation, multiplying the
  Killing vectors.

 The Killing vectors
  $(V^\sca)^i{}_j\partial_{y^\sca}$ satisfy  the  $so(6)\simeq su(4)$
commutation relations  (\ref{su4com}) and  may be written as
\begin{equation}
(V^\sca)^i{}_j\partial_{y^\sca}
=\frac{1}{4}(\gamma^{\apr\bpr})^i{}_jV^{\apr\bpr}
+\frac{\rm i}{2}(\gamma^\apr)^i{}_j V^\apr      \ ,
\end{equation}
where  $V^\apr$ and $V^{\apr\bpr}$
correspond to the  5 translations and  $SO(5)$
rotations respectively
and  are given by (cf. \rf{uuuu})
\begin{eqnarray}
&&
V^\apr
=\Bigl[|y|\cot |y| (\delta^{\apr\sca}-n^\apr n^\sca)
+n^\apr n^\sca\Bigr]\partial_{y^\sca}  \ ,
\\
&&
V^{\apr\bpr}=y^\apr \partial_{y^\bpr}-y^\bpr \partial_{y^\apr}  \ .
\end{eqnarray}
Here $\delta^{\apr\sca}$ is  Kronecker delta symbol and we
use the conventions: $y^\sca=\delta_\apr^\sca y^\apr$,\
$n^\sca=\delta_\apr^\sca n^\apr$, $n^\sca=n_\sca$.
In  these coordinates the  $S^5$ metric
tensor  has the form
\begin{equation}
G_{\sca\scb}= e_\sca^\apr e_\scb^\apr\,,   \ \
\qquad
e_\sca^\apr=\frac{\sin |y|}{|y|}(\delta_\sca^\apr-n_\sca n^\apr)
+n_\sca n^\apr    \ .
\end{equation}
Note that while deriving (\ref{actkin3}) we use the relation 
$(U^\dagger \gamma^\apr U)^i{}_j=-2{\rm i}e_\sca^\apr(V^\sca)^i{}_j$.

The Lagrangian \rf{actkin3},\rf{actwz3}
can be put into the  manifestly $SU(4)$ 
invariant form by  changing the coordinates from
 $\p, y^{\apr}$ to the  Cartesian coordinates  $Y^M$\ ($M=1,...,6$):
\be
Y^\apr = e^\phi \sin|y|\ n^\apr\,, 
\qquad
Y^6 = e^\phi \cos|y|
\ , \quad Y^2 = Y^M Y^M = |Y|^2 = e^{2 \phi} \ . 
\la{coch}
\ee
In terms of the new coordinates the  superstring Lagrangian 
$ {\cal L} = {\cal L}_{kin} + {\cal L}_{WZ}$
takes then the following more 
transparent manifestly $SU(4)$ invariant
form: 
$$
{\cal L}_{kin}
= -
\sqrt{g}g^{\vm\vn}\Bigl[
Y^2(\partial_\vm x^+ \partial_\vn x ^-
+ \partial_\vm x\partial_\vn\bar{x})
+ \frac{1}{2Y^2}D_\vm Y^M D_\vn Y^M\Bigr]
$$
\be\la{goki}
- \ \frac{{\rm i}}{2} \sqrt{g}g^{\vm\vn}
Y^2\partial_\vm x^+
\Bigl[\theta^i\partial_\vn \theta_i
+\theta_i\partial_\vn \theta^i
+\eta^i\partial_\vn \eta_i
+\eta_i\partial_\vn \eta^i 
+ {\rm i}  Y^2\partial_\vn x^+(\eta^2)^2\Bigr]\ ,
%\label{actkin4n}
\ee
\begin{equation}\label{gowz}
{\cal L}_{WZ}
=\epsilon^{\vm\vn}
|Y|\partial_\vm x^+ \eta^i Y^M\rho_{ij}^M 
\bigg(\partial_\vn\theta^j-{\rm i}\sqrt{2}|Y| \eta^j
\partial_\vn x\bigg)+h.c.\ , 
\end{equation}
where 
\be
DY^M = dY^M -2{\rm i}\eta_i (R^M)^i{}_j\eta^j Y^2 dx^+ \ . 
\la{codi}
\ee
The 6 matrices  $\rho_{ij}^M$  are the $SO(6)$   $\gamma$   matrices 
in the 
chiral representation. The usual $SO(6)$  Dirac 
 matrices  can be  expressed 
in terms of $\rho_{ij}^M$ as follows
\be\label{usgam}
\gamma^M
=\left(\begin{array}{cc}
 0   & (\rho^M)^{ij} 
 \\
 \rho_{ij}^M & 0
 \end{array}
 \right)\,,
 \qquad
 (\rho^M)^{il}\rho_{lj}^N + (\rho^N)^{il}\rho_{lj}^M
 =2\delta^{MN}\delta_j^i\,,
 \qquad
 \rho_{ij}^M =- \rho_{ji}^M\,,
 \ee
where $(\rho^M)^{ij}= - (\rho_{ij}^{M})^*$. In deriving (\ref{gowz})
we used the following
representation for $C^\prime$ and $SO(5)$ $\gamma$ matrices in terms 
of the 
$\rho$ matrices
\be
(\gamma^\apr)^i{}_j = {\rm i}(\rho^\apr)^{il}\rho_{lj}^6\,,
\qquad
C_{ij}^\prime =\rho_{ij}^6 \ , 
\ee
implying an interesting relation
\be\label{intrel}
e^\phi C_{ij}^U  = \rho_{ij}^M Y^M\,.
\ee
The matrices $(R^M)^i{}_j$ in the covariant derivative 
\rf{codi} are defined as follows  ($M,\! N,K,L\! =1,...,6$) 
\be
(R^M)^i{}_j = \frac{1}{4}(\rho^{KL})^i{}_j (R^M)^{KL}\,,
\la{fffr}
\ee
where
\be
(R^M)^{KL} = Y^K \delta^{LM} - Y^L \delta^{KM}\,,
\qquad
(\rho^{KL})^i{}_j \equiv \frac{1}{2}(\rho^K)^{il}\rho_{lj}^L
-(K\leftrightarrow L)\,.
\ee
Note that $(R^M)^i{}_j\partial_{Y^M}$ satisfy
the  $so(6)\simeq su(4)$
commutation relations  (\ref{su4com}).
In contrast to $V^\sca$ which are complicated functions of $y^\sca$
the  matrices  $R^M$ take simpler form. 

Note that in terms of the 6  Cartesian coordinates
 $Y^M$  the metric of \adss takes the ``4+6'' 
form: $$ds^2 =
Y^2 dx^a dx^a +  Y^{-2} dY^M dY^M\ .$$ Similar
choice  of the  bosonic part of superstring  
coordinates was used, e.g., in 
\ci{KT,pes}.
The advantage of the resulting action is a more transparent structure of the WZ
term \rf{gowz}.

The above action \rf{goki},\rf{gowz} 
can be transformed into the equivalent 
form  corresponding to the choice of the 
 conformally flat coordinates in \adss, i.e. 
 ($ Y^M \to { Z^M \ov Z^2}$)
 $$  ds^2 =
{ 1 \ov Z^2} (  dx^a dx^a +   dZ^M dZ^M )\ .  $$ 
If we   start again with (\ref{actkin3}),(\ref{actwz3}) 
and  introduce  (cf. (\ref{coch}))
\be 
Z^\apr = e^{-\phi} \sin|y| \ n^\apr\,,
\qquad
Z^6 = e^{-\phi} \cos|y| \ ,
 \quad Z^2 = Z^M Z^M = |Z|^2 = e^{-2 \phi} \ ,  
\ee
then we finish with  (cf. \rf{goki},\rf{gowz})
$$
{\cal L}_{kin}
= -
\sqrt{g}g^{\vm\vn}Z^{-2}\Bigl[
\partial_\vm x^+ \partial_\vn x ^-
+ \partial_\vm x\partial_\vn\bar{x}
+ \frac{1}{2}D_\vm Z^M D_\vn Z^M\Bigr]
$$
\begin{equation}
- \ \frac{{\rm i}}{2} \sqrt{g}g^{\vm\vn}
Z^{-2}\partial_\vm x^+
\Bigl[\theta^i\partial_\vn \theta_i
+\theta_i\partial_\vn \theta^i
+\eta^i\partial_\vn \eta_i
+\eta_i\partial_\vn \eta^i 
+{\rm i}  Z^{-2}\partial_\vn x^+(\eta^2)^2\Bigr]\ ,
\label{gokii}
\end{equation}
\begin{equation}\label{actwz5n}
{\cal L}_{WZ}
=\epsilon^{\vm\vn}
|Z|^{-3}\partial_\vm x^+ \eta^i \rho_{ij}^M Z^M
\Bigl(\partial_\vn\theta^j-{\rm i}\sqrt{2}|Z|^{-1} \eta^j
\partial_\vn x\Bigr)+h.c. \ , 
\la{gowzz}
\end{equation}
where $Z^{-2}\equiv (Z^2)^{-1}$  and (cf. \rf{codi},\rf{fffr})  
\be
DZ^M = dZ^M -2{\rm i}\eta_i (R^M)^i{}_j\eta^j Z^{-2}dx^+ \ ,
  \quad \  
  R^M = -\frac{1}{2}\rho^{ML} Z^L \,.
 \la{codii}
\ee
All  other  notation are the same as  above.
One can obtain  (\ref{gokii}), (\ref{gowzz})
directly from \rf{goki},\rf{gowz} by  making  the 
inversion  $Y^M \rightarrow Z^M/Z^2$ and 
 taking into account the relation $R^M Z^M=0$. 

\bigskip

In  this Section we have discussed
the light-cone action in  the Killing  parametrization  of
superspace. In order to get the light-cone gauge action in
the Wess-Zumino parametrization
one needs to make the following  redefinitions
in (\ref{actkin2}), (\ref{actwz2}) (cf. (\ref{tilthe}), (\ref{tildthe}))
\be
\theta^i \rightarrow  (U^{-1})^i{}_j\theta^i\,, \ \ \
\qquad
\theta_i \rightarrow \theta_j U^j{}_i
\ee
\be
\eta^i \rightarrow \sqrt{2}e^\phi (U^{-1})^i{}_j\eta^j\,,
\qquad
\eta_i \rightarrow \sqrt{2}e^\phi\eta_j U^j{}_i    \ .
\ee
In addition we change sign  of 4-d coordinates
 $x^a\rightarrow -x ^a$. 
The fermionic derivatives $\partial_\mu $
will  then   get  the generalized
connection $\Omega_\mu = \del_\mu U U^{-1} $ (\ref{gencon}) contributions, i.e.
  become the covariant derivatives
${\cal D}_\mu$ (see  (\ref{covder})). The action in terms of these new
variables was   presented in (\ref{ff}),(\ref{qua}) in the Introduction.

Finally, let us note that  our
 results for  the \adss  space
can be generalized to the case of $AdS_3\times S^3$
in a rather straightforward way. To get the light-cone   gauge
action for this case  one  could use the $\kappa$
invariant action of Ref.\cite{adstri}  and then
apply the same procedure of light-cone
splitting and gauge fixing as  developed in this paper.
However,  our  light-cone gauge action
 is already written in the form which
 allows a straightforward
generalization to the case of $AdS_3\times S^3$:
one is just to do a     dimensional
reduction.
Let us discuss the $AdS_3\times S^3$ Lagrangian
using  for definiteness  the WZ parametrization
where the action has the form given by (\ref{oom}).
 To get
 the ${\cal L}_B$ and
${\cal L}_F^{(2)}$ terms  in the
$AdS_3\times S^3$   case  we are  to set
$x=\bar{x}=0$ in  (\ref{bos}) and (\ref{ff})
  and also to assume that
 the fermionic variables $\theta$ and $\eta$
now   transform in the  fundamental representation of $SU(2)$ (i.e.  the
indices $i,j$ take values $1,2$).
 The matrix $\Csp_{ij}$ is then  given
by $\Csp= h \sigma_2$,\ $|h|=1$. The matrices $(\gamma^\apr)^i{}_j$,
$\apr=1,2,3$ are now  $SO(3)$ Dirac gamma matrices and the matrix
$U(y)$ takes the same form as in (\ref{uuu}), (\ref{uuuu}).
The quartic part of the  Lagrangian ${\cal L}_F^{(4)}$
in \rf{qua} simplifies to\foot{To transform (\ref{qua}) to this form we use the completeness relation
for $SO(3)$ gamma matrices
$
(\gamma^\apr)^i{}_j(\gamma^\apr)^k{}_l
=-\delta_j^i\delta_l^k + 2\delta_l^i\delta_j^k  \ .
$
}
\begin{equation}
\label{qua33}
{\cal L}^{(4)}_{F} =  2\sqrt{g} g^{\vm\vn}
 e^{4\phi}\partial_\vm x^+ \partial_\vn x^+  (\eta^i\eta_i)^2   \ .
\end{equation}

%%%%%%%%%%%%%%%%%%%%%%%%%%%%%%%%%%%%%%%%%%%%%%%%%%%%%%%%%%%%%%
%%%%%%%%%%%%%%%%%%%%%%%%%%%%%%%%%%%%%%%%%%%%%%%%%
\section*{Acknowledgments}
%%%%%%%%%%%%%%%%%%%%%%%%%%%%%%%%%%%%%%%%%%%%%%%%%%

We are grateful to A.M. Polyakov and C. Thorn for
stimulating discussions  and  suggestions, and to S. Frolov 
for useful comments.
  This work   was  supported in part by
the DOE grant DOE/ER/01545 and the INTAS project 991590.
 R.R.M. is also supported by the RFBR Grant No 99-02-17916.
A.A.T. would like to acknowledge also
the support of the
EC TMR grant ERBFMRX-CT96-0045
and PPARC SPG grant  PPA/G/S/1998/00613.

%%%%%%%%%%%%%%%%%%%%%%%%%%%%%%%%%%%%%%%%%%%%%%%%
\setcounter{section}{0}
\setcounter{subsection}{0}

\appendix{$psu(2,2|4)$ superalgebra:  $so(4,1)\oplus so(5)$,
$so(3,1)\oplus su(4)$  and
light-cone bases}
%%%%%%%%%%%%%%%%%%%%%%%%%%%%%%%%%%%%%%%%%%%%%%%%%%%

Commutation relations of $psu(2,2|4)$ superalgebra in
$so(4,1)\oplus so(5)$ basis  were given  in \cite{MT}. This basis
is most adequate for  finding   the covariant action in
  $AdS_5\times  S^5$ space \cite{MT}   which is the direct analogue
 of
the GS action in  flat space.
 To develop the light-cone
formulation it is convenient,  however,
 to make a  transformation to the basis in
which the supercharges are diagonal with respect to
the generators $J^{+-}$,
$D$, $J^{x\bar{x}}$ (see \rf{gene})
and  belong to the fundamental representation of
$su(4)$. This basis we shall call light-cone basis.

We shall     find the transformation to the
light-cone basis at the
level of the algebra,  and this will
allow us to  find the Cartan 1-forms and the action
in the form corresponding to the
 light-cone basis.
It is convenient to first
 make the transformation to the  intermediate
$so(3,1)\oplus su(4)$ basis and only then to the light-cone basis.
A bonus of this procedure is that this intermediate form will
allow us to find as a by-product
another interesting  version of the  $\kappa$-symmetry  gauge fixed
action (see Appendix C).

 We start with the commutation
relations of $psu(2,2|4)$ superalgebra in $so(4,1)\oplus so(5)$ basis
given in \ci{MT}
\begin{equation}\label{cmr1}
[\hat{P}_\aA,\hat{P}_\aB]=\hat{J}_{\aA\aB}\,,
\qquad
[P_\apr, P_\bpr]=-J_{\apr\bpr}\,,
\end{equation}
\begin{equation}\label{cmr2}
[\hat{J}^{\aA\aB},\hat{J}^{\aC\aE}]
=\eta^{\aB\aC}\hat{J}^{\aA\aE}+3 \hbox{ terms},
\qquad
[J^{\apr\bpr},J^{\cpr\epr}]=\eta^{\bpr\cpr}J^{\apr\epr}+3 \hbox{ terms} \ ,
\end{equation}
\be
[Q_{_I},\hat{P}_\aA]
=-\frac{{\rm i}}{2}\epsilon_{_{IJ}}Q_{_J}\gamma_\aA \,,
\qquad
[Q_{_I},\hat{J}_{\aA\aB}]=-\frac{1}{2} Q_{_I}\gamma_{\aA\aB}\,,
\ee
\be
[Q_{_I},P_\apr]
=\frac{1}{2}\epsilon_{_{IJ}} Q_{_J}\gamma_\apr\,,
\qquad
[Q_{_I},J_{\apr\bpr}]=-\frac{1}{2}Q_{_I}\gamma_{\apr\bpr} \,,
\ee
\begin{eqnarray}
\{Q_{\alpha \alpr I}, Q_{\beta \bepr J}\}
&=&\delta_{_{IJ}}
[-2{\rm i}C_{\alpr\bepr}^\prime (C\gamma^\aA)_{\alpha\beta} \hat{P}_\aA
+2C_{\alpha\beta}(C^\prime\gamma^\apr)_{\alpr\bepr}P_\apr]
\nonumber\\
&+&\epsilon_{_{IJ}}
[C_{\alpr\bepr}^\prime (C\gamma^{\aA\aB})_{\alpha\beta} \hat{J}_{\aA\aB}
-C_{\alpha\beta}(C^\prime\gamma^{\apr\bpr})_{\alpr\bepr}J_{\apr\bpr}] \ .
\end{eqnarray}
Unless otherwise  specified,
  we use the notation
$Q^{^I}$ for $Q^{I\alpha\alpr}$ and $Q_{_I}$ for $Q_{I\alpha\alpr}$, 
where $
Q_{I\alpha\alpr}\equiv
Q^{J\beta\bepr}\delta_{JI}C_{\beta\alpha}C_{\bepr\alpr}^\prime
. $
Hermitean conjugation rules in this basis are
\be
\hat{P}_\aA^\dagger=- \hat{P}_\aA\,,\ \ \ \ \
P_\apr^\dagger= - P_\apr \ , \ \ \
\ \
\hat{J}_{\aA\aB}^\dagger=- \hat{J}_{\aA\aB}\,,\ \ \ \ \
\ \
J_{\apr\bpr}^\dagger=- J_{\apr\bpr}
\ ,\ee \be  \ \
(Q^{I\beta\alpr})^\dagger (\gamma^0)^\beta_\alpha
= - Q^{I\beta\bepr}C_{\beta\alpha}C_{\bepr\alpr}^\prime\ .
\ee

Let us first  transform the  bosonic generators into
the conformal algebra basis.
To this end we introduce the
Poincar\'e translations $P^a$, the conformal boosts $K^a$
and the dilatation $D$ by
\be
P^a = \hat{P}^a +\hat{J}^{4a}
\ , \ \ \ \
K^a=\frac{1}{2}(-\hat{P}^a +\hat{J}^{4a})
\ ,  \ \ \ \
D=-\hat{P}^4 \ .  \label{dp4}
\ee
Making use of the commutation relations (\ref{cmr1}),(\ref{cmr2}) it is easy
to check that these  generators satisfy
the commutation relations given in
(\ref{comrel1}),(\ref{comrel2}).

Next,   we introduce the new ``charged"  super-generators
\be
Q^q\equiv \frac{1}{\sqrt{2}}(Q^1+{\rm i}Q^2)   \ ,
\qquad  \ \ \
Q^{\bar{q}}
\equiv \frac{1}{\sqrt{2}}(Q^1-{\rm i}Q^2)
\ . \ee
We shall use the simplified notation
\begin{equation}\label{intsc2}
Q^{\alpha\alpr}\equiv -Q^{q\alpha\alpr}\,,
\qquad
Q_{\alpha\alpr}\equiv Q_{q\alpha\alpr}\,.
\end{equation}
Then   the non-vanishing values of
$\delta_{IJ}$  ($\epsilon_{IJ}$, $\epsilon_{12}=1$) become replaced by
$\delta_{q\bar{q}}=1$ ($\epsilon_{q\bar{q}}={\rm i}$) and
the Majorana condition  takes  the form  $
(Q^{\beta\alpr})^\dagger(\gamma^0)^\beta_\alpha
=Q_{\alpha\alpr}.
$
The  commutators  have the form
\be
[Q^{\alpha\alpr},\hat{P}^\aA]
=-\frac{1}{2}(\gamma^\aA Q)^{\alpha\alpr}\,,
\qquad
[Q^{\alpha\alpr},\hat{J}^{\aA\aB}]
=\frac{1}{2}(\gamma^{\aA\aB}Q)^{\alpha\alpr}
\ , \ee
\be
[Q_{\alpha\alpr},\hat{P}^\aA]
=\frac{1}{2}(Q\gamma^\aA)_{\alpha\alpr}\,,
\qquad
[Q_{\alpha\alpr},\hat{J}^{\aA\aB}]
=-\frac{1}{2}(Q\gamma^{\aA\aB})_{\alpha\alpr}
\ , \ee
\begin{equation}\label{qps1}
[Q^{\alpha\alpr},P^\apr]
=-\frac{{\rm i}}{2}(\gamma^\apr Q)^{\alpha\alpr}\ ,
\qquad
[Q^{\alpha\alpr},J^{\apr\bpr}]
=\frac{1}{2}(\gamma^{\apr\bpr}Q)^{\alpha\alpr}  \ ,
\end{equation}
\begin{equation}\label{qps2}
[Q_{\alpha\alpr},P^\apr]
=\frac{{\rm i}}{2}(Q\gamma^\apr)_{\alpha\alpr}  \ ,
\qquad
[Q_{\alpha\alpr},J^{\apr\bpr}]
=-\frac{1}{2}(Q\gamma^{\apr\bpr})_{\alpha\alpr} \ ,
\end{equation}
while the anti-commutators transform into the form
\be
\{Q^{\alpha\alpr},Q_{\beta\bepr}\}=[2{\rm i}(\gamma^\aA)^\alpha_\beta
\hat{P}^\aA
+(\gamma^{\aA\aB})^\alpha_\beta \hat{J}^{\aA\aB}]\delta_\bepr^\alpr
-4{\rm i}\delta_\beta^\alpha J^\alpr{}_\bepr  \ ,
\ee
where we use the notation
\begin{equation}\label{su4def}
J^\alpr{}_\bepr \equiv
-\frac{{\rm i}}{2}(\gamma^\apr)^\alpr{}_\bepr P^\apr
+\frac{1}{4}(\gamma^{\apr\bpr})^\alpr{}_\bepr J^{\apr\bpr}   \ .
\end{equation}
Starting with   the commutation relations for $P^\apr$ and  $J^{\apr\bpr}$ and
applying  various Fierz identities one proves that $J^\alpr{}_\bepr$
\ ($J^i{}_j^\dagger=J^j{}_i$) satisfy
the commutation relations of $su(4)$ algebra.

Using the commutators (\ref{qps1}),(\ref{qps2}) and (\ref{su4def}) and
completeness relation for Dirac  matrices one proves that
\be
[Q_{\alpha\alpr},J^\bepr{}_\gapr]=\delta_\alpr^\bepr Q_{\alpha\gapr}
-\frac{1}{4}\delta_\gapr^\bepr Q_{\alpha\alpr}\,,
\qquad
[Q^{\alpha\alpr},J^\bepr{}_\gapr]=-\delta^\alpr_\gapr Q^{\alpha\bepr}
+\frac{1}{4}\delta^\bepr_\gapr Q^{\alpha\alpr}
\ . \ee
This demonstrates  that supercharges transform in the  fundamental
representations of $su(4)$.

In what follows we will use the following decomposition of $so(4,1)$
 Dirac  and  charge conjugation  matrices in  the $sl(2)$ basis
\begin{equation}\label{gamdec}
(\gamma^a)^\alpha{}_\beta
=\left(
\begin{array}{cc}
0 & (\sigma^a)^{\sfa\dsfb  }
\\
\bar{\sigma}^a_{\dsfa \sfb}  & 0
\end{array}
\right)\,,
\qquad
\gamma^{4}
=\left(
\begin{array}{cc}
1 & 0
\\
0  & -1
\end{array}
\right)\,,
\qquad
C_{\alpha\beta}
=\left(
\begin{array}{cc}
\epsilon_{\sfa\sfb} & 0
\\
0  & \epsilon^{\dsfa \dsfb  }
\end{array}
\right)\,,
\end{equation}
where  the matrices $(\sigma^a)^{\sfa\dsfa}$,
$(\bar{\sigma}^a)_{\dsfa\sfa}$ are related to Pauli matrices
in the standard way
\be
\sigma^a =(1,\sigma^1,\sigma^2,\sigma^3)\,,
\qquad
\bar{\sigma}^a =(-1,\sigma^1,\sigma^2,\sigma^3)\, . \ee
Note that $\sigma_{\sfa\dsfa}^a=\bar{\sigma}_{\dsfa\sfa}^a$,
$\bar{\sigma}_{\dsfa\sfb}^a=\sigma_{\sfa\dsfb}^{a*}$ where
$\sigma_{\sfa\dsfa}^a\equiv
(\sigma^a)^{\sfb\dsfb}\epsilon_{\sfb\sfa}\epsilon_{\dsfb\dsfa}$.
We use the following conventions for the $sl(2)$ indices: \ 
$
\epsilon_{12}=\epsilon^{12}=-\epsilon_{\dot{1}\dot{2}}
=-\epsilon^{\dot{1}\dot{2}}=1    \ ,
$
\begin{equation}\label{rlrul}
\psi^\sfa=\epsilon^{\sfa\sfb}\psi_\sfb  \ ,
\quad
\psi_\sfa=\psi^\sfb\epsilon_{\sfb\sfa}    \ ,
\qquad
\psi^{\dsfa }=\epsilon^{\dsfa \dsfb  }\psi_{\dsfb  }\ ,
\quad
\psi_{\dsfa }=\psi^{\dsfb  }\epsilon_{\dsfb  \dsfa } \ .
\end{equation}
We then decompose the supercharges in the $sl(2)\oplus su(4)$ basis
\be
Q^{\alpha i}
=\left(
\begin{array}{c}
2{\rm i}v^{-1}Q^{\sfa  i}
\\[7pt]
2vS_{\dsfa }^i
\end{array}\right)  \ ,
\qquad    \
Q_{\alpha i}=(2vS_{\sfa i}, -2{\rm i}v^{-1}Q_i^{\dsfa })   \ ,   \ \ \
\ \  v\equiv 2^{1/4} \ .
\ee
In terms of these  new supercharges  the commutation relations take the form
\be
[D,Q^{\sfa i}]=-\frac{1}{2}Q^{\sfa i}\,,
\qquad
[D,S_i^\sfa]=\frac{1}{2}S_i^\sfa   \ ,
\ee
\be
[S^\sfa_i,P^a]
=\frac{{\rm i}}{\sqrt{2}}(\sigma^a)^{\sfa\dsfa }Q_{\dsfa i}\,,
\qquad
[S_{\dsfa }^i,P^a]
=-\frac{{\rm i}}{\sqrt{2}}(\bar{\sigma}^a)_{\dsfa \sfa}Q^{\sfa  i}
\ , \ee
\be
[Q^{\sfa  i},K^a]
=-\frac{{\rm i}}{\sqrt{2}}(\sigma^a)^{\sfa\dsfa }S_{\dsfa }^i\,,
\qquad
[Q_{\dsfa i},K^a]
=\frac{{\rm i}}{\sqrt{2}}(\bar{\sigma}^a)_{\dsfa \sfa}S^\sfa_i
\ , \ee
\be
\{Q^{\sfa  i},Q^{\dsfb  }_j\}
=\frac{{\rm i}}{\sqrt{2}}\sigma_a^{\sfa\dsfb  }P^a\delta_j^i\,,
\qquad
\{S_j^{\sfa},S^{\dsfb  i}\}
=-\frac{{\rm i}}{\sqrt{2}}\sigma_a^{\sfa\dsfb  }K^a\delta_j^i
\ , \ee
\be
[Q^{\sfa i},J^{ab}]
=\frac{1}{2}(\sigma^{ab})^\sfa{}_\sfb Q^{\sfb i}
\ , \ee
\be
\{Q^{\sfa  i},S^\sfb_j\}
=(\frac{1}{2}\epsilon^{\sfa\sfb}D
+\frac{1}{4}\sigma_{ab}^{\sfa\sfb}J^{ab})\delta_j^i
+\epsilon^{\sfa\sfb}J^i{}_j
\ , \ee
where $
(\sigma^{ab})^{\sfa\sfb}= \epsilon^{\sfb\sfc}(\sigma^{ab})^\sfa{}_\sfc\ ,
\ \ \
(\sigma^{ab})^\sfa{}_\sfb\equiv\frac{1}{2}
(\sigma^a)^{\sfa\dsfc}(\bar{\sigma}^b)_{\dsfc\sfb}
-(a\leftrightarrow b)   \ .
$
Hermitean conjugation rules of the supercharges  are
\begin{equation}\label{herconrul5}
Q^{i \sfa\dagger} =Q^{\dsfa }_i\,,
\qquad
Q_\sfa^{i\dagger} = - Q_{\dsfa i}  \ ,
\end{equation}
and the same  for $S$ supercharges.
The spinor $sl(2)$ indices
$\sfa,\sfb $ are raised and lowered as
in (\ref{rlrul}). From these commutation
relations we learn that $Q^{\sfa i}$, $Q_i^{\dsfa }$
may be interpreted as the  supercharges of the super
Poincar\'e subalgebra while
$S^\sfa_i$, $S^{\dsfa i}$ are the conformal supercharges.

This finishes the description of the $so(3,1) \oplus su(4)$
basis.
We are now ready to introduce the light-cone basis. The transformation of
the
bosonic generators is implied by the light-cone decomposition
of the coordinates (\ref{lcbas})
and is  given by \rf{gene}.  The  transformation  of  supercharges
amounts to attaching the signs  $+$ and $-$
 which will  show explicitly their
$J^{+-}$ charges. The corresponding
supercharges are defined by
\be
Q^{1i}\equiv -Q^{-i}\,,
\qquad
Q^{2i}\equiv Q^{+i}\,,
\qquad
Q^{\dot{1}}_i\equiv -Q^-_i\,,
\qquad
Q^{\dot{2}}_i\equiv Q^+_i\,,
\ee
\be
S^1_i\equiv S^-_i\,,
\qquad
S^2_i\equiv -S^+_i\,,
\qquad
S^{\dot{1}i}\equiv S^{-i}\,,
\qquad
S^{\dot{2}i}\equiv -S^{+i}\, .
\ee
Choice of signs in these definitions is a  matter of convention.
Hermitean conjugation rules (\ref{herconrul5}) lead then to
the conjugation
rules given in (\ref{herconrul3}).

%%%%%%%%%%%%%%%%%%%%%%%%%%%%%%%%%%%%%%%%%%%%%%%%%%%
\appendix{Cartan forms
in $so(3,1)\oplus su(4)$ and light cone bases}
%%%%%%%%%%%%%%%%%%%%%%%%%%%%%%%%%%%%%%%%%%%%%%%%%%%%%%%%%

The kinetic term  of the \adss GS  action
and the 3-form  in its WZ term  have  the following
form  in  the $so(4,1)\oplus so(5)$ basis \ci{MT}
\begin{equation}\label{lkin2}
{\cal L}_{kin}=-\frac{1}{2}\sqrt{g}g^{\vm\vn}
(\hat{L}_{\vm}^\aA \hat{L}_{\vn}^\aA
+L_{\vm}^\apr L_{\vn}^\apr) \ ,
\end{equation}
\begin{equation}\label{wzt}
{\cal H}=s^{IJ}\hat{L}^\aA \bar{L}^I\gamma^\aA L^J
+{\rm i}s^{IJ}L^\apr \bar{L}^J\gamma^\apr L^J  \ .
\end{equation}
They  are  expressed  in terms of the Cartan 1-forms
defined in the $so(4,1)\oplus so(5)$ basis   by
\be
G^{-1}dG=(G^{-1}dG)_{bos}
+L^{I\alpha \alpr}Q_{I\alpha\alpr}
\ , \ee
where the restriction to  the  bosonic part is
\begin{equation}\label{cf1}
(G^{-1}dG)_{bos}
=\hat{L}^\aA\hat{P}^\aA+\frac{1}{2}\hat{L}^{AB}\hat{J}^{\aA\aB}
+L^\apr P^\apr+\frac{1}{2} L^{\apr\bpr} J^{\apr\bpr}\ .
\end{equation}
The transformation of the $psu(2,2|4)$ algebra into
the light-cone basis described  in
Appendix A  allows us to find the  corresponding Cartan
1-forms  and thus to write down  the GS
 action in  the light-cone basis.

We first consider the
 $so(3,1)\oplus su(4)$ basis  and  define  the
 bosonic (even) Cartan forms by
\begin{equation}\label{cf2}
(G^{-1}dG)_{bos}
=L_\ssmP^aP^a+L_\ssmK^a K^a +L_\ssmD D + \frac{1}{2}L^{ab}J^{ab}
+L^\alpr{}_\bepr J^\bepr{}_\alpr      \ .
\end{equation}
Comparing this  with (\ref{cf1}) and using (\ref{dp4}),
(\ref{su4def}) we get
\be
\hat{L}^a = L_\ssmP^a - \frac{1}{2}L_\ssmK^a
\ , \ \ \ \
\hat{L}^{4a}=L_\ssmP^a+\frac{1}{2}L_\ssmK^a
\ , \ \ \ \
\hat{L}^4=-L_\ssmD\ ,
\ee
\be
\label{lijlab}
L^i{}_j=\frac{\rm i}{2}(\gamma^\apr)^i{}_j L^\apr
-\frac{1}{4}(\gamma^{\apr\bpr})^i{}_j L^{\apr\bpr}   \ .
\ee
Using  these relations in the expression for
the kinetic term (\ref{lkin2})
gives the action (\ref{lkin1}).

Now let us consider  the fermionic  1-forms. They satisfy
hermitean conjugation rule
\be
(L^{I\beta\alpr})^\dagger (\gamma^0)^\beta_\alpha
= L^{I\beta\bepr}C_{\beta\alpha}C_{\bepr\alpr}^\prime.
\ee
and we use the notation
$L_{I\alpha\alpr}\equiv
L^{J\beta\bepr}\delta_{JI}C_{\beta\alpha}C_{\bepr\alpr}^\prime$.
Let us   define
\begin{equation}\label{lq}
L^q\equiv \frac{1}{\sqrt{2}}(L^1+{\rm i}L^2)   \ ,
\qquad  \ \
L^{\bar{q}}\equiv \frac{1}{\sqrt{2}}(L^1-{\rm i}L^2) \ ,
\end{equation}
introduce the notation
$L^{\alpha i}= L^{q\alpha i}$, $L_{\alpha i}= L_{q\alpha i}$
and use  the following decomposition into
$sl(2)\oplus su(4)$ Cartan 1-forms
\be
L^{\alpha i}
=\frac{1}{2} \left(
\begin{array}{c}
v^{-1}L_\ssmS^{\sfa  i}
\\[7pt]
{\rm i}v L_{\ssmQ\,\dsfa }^i
\end{array}\right) \ , \ \ \ \
\qquad
L_{\alpha i}=\frac{1}{2} (- {\rm i}v L_{\ssmQ\,\sfa i},\,
v^{-1}L_{\ssmS i}^{\dsfa })
\ . \ee
Hermitean conjugation rules for the new Cartan 1-forms  then
take the same form
as in (\ref{herconrul5}).
The light-cone frame Cartan 1-forms are defined
by
\begin{eqnarray}
L_{\ssmQ i}^1=-L_{\ssmQ i}^-\,,
\quad
L_{\ssmQ i}^2=-L_{\ssmQ i}^+\,,
\qquad \ \
L_\ssmQ^{\dot{1} i}= -L_\ssmQ^{-i}\,,
\quad
L_\ssmQ^{\dot{2} i}=-L_\ssmQ^{+i}\,,
\\
\label{hhh}
L_\ssmS^{1i} = L_\ssmS^{-i}\,,
\quad
L_\ssmS^{2i} = L_\ssmS^{+i}\,,
\qquad \ \
L_{\ssmS i}^{\dot{1}}=L_{\ssmS i}^-\,,
\quad
L_{\ssmS i}^{\dot{2}} = L_{\ssmS i}^+\, .
\end{eqnarray}
These   relations imply
\begin{eqnarray}
L^{I\alpha i}Q_{I\alpha i}
&=&L^{\alpha i}Q_{\alpha i}
-L_{\alpha i}Q^{\alpha i}
\\ \label{2l}
&=&
 L_{\ssmQ i}^\sfa Q_\sfa^i
-L_\ssmQ^{\dsfa i} Q_{{\dsfa }i}
+L_\ssmS^{\sfa i}S_{\sfa i}
-L_{\ssmS i}^{\dsfa }S_{\dsfa }^i
\\  \nonumber
&=&
  L_\ssmQ^{+i} Q_i^-
+ L_\ssmQ^{-i} Q_i^+
+ L_{\ssmQ i}^+ Q^{- i}
+ L_{\ssmQ i}^-Q^{+ i}
\\
\label{4l}&& + \
 L_\ssmS^{-i}S_i^+
+L_\ssmS^{+i}S_i^-
+L_{\ssmS i}^-S^{+i}
+L_{\ssmS i}^+S^{-i}   \ .
\end{eqnarray}
The representation (\ref{2l}) corresponds to    the
$sl(2)\oplus su(4)$ basis while (\ref{4l}) --  to the  light-cone basis.

Using the relation between the Cartan 1-forms in
(\ref{lq})--(\ref{hhh}) we are ready to consider
the  decomposition of the WZ  3-form
 (\ref{wzt}).  We start with
 the $AdS_5$  contribution which is given by the first
term in r.h.s of  (\ref{wzt}). Taking into account that
$\bar{L}^I=L^IC\Csp$ and eq. (\ref{lq}) we can rewrite
the  $AdS_5$ contribution in
terms of the ``charged"  Cartan forms $L^q$, $L^{\bar{q}}$
\begin{equation}\label{dec1}
{\cal H}_{AdS_5}={\cal H}_{AdS_5}^q+{\cal H}_{AdS_5}^{\bar{q}}  \ ,
\end{equation}
\be
{\cal H}_{AdS_5}^q\equiv
\hat{L}^\aA
L^{q\alpha i}(C\gamma^\aA)_{\alpha\beta} \Csp_{ij} L^{q\beta j}  \ ,
\qquad                     \ \
{\cal H}_{AdS_5}^{\bar{q}}\equiv
\hat{L}^\aA
L^{\bar{q}\alpha i}(C\gamma^\aA)_{\alpha\beta} 
\Csp_{ij} L^{\bar{q}\beta j} \ .
\ee
Since ${\rm i}{\cal H}_{AdS_5}^{\bar{q}}$ is hermitean
conjugate to  ${\rm i}{\cal H}_{AdS_5}^q$ we restrict our
attention to  decomposition of the first term.
We get
\begin{eqnarray}
{\cal H}_{AdS_5}^q
&=&\hat{L}^a L^{q\alpha i}(C \gamma^a)_{\alpha\beta}\Csp_{ij} L^{q\beta j}
-L_\ssmD L^{q\alpha i}(C\gamma^4)_{\alpha\beta} \Csp_{ij} L^{q\beta j}
\\ \nonumber
\label{3line}
&=&\frac{{\rm i}}{2}\hat{L}^a
L_{\ssmS \sfa}^i\Csp_{ij}(\sigma^a)^{\sfa\dsfb} L_{\ssmQ \dsfb  }^j
+\frac{1}{4}L_\ssmD
(\frac{1}{\sqrt{2}}L_\ssmS^{\sfa i}\Csp_{ij}L_{\ssmS \sfa}^j
+\sqrt{2}L_\ssmQ^{\dsfa i}\Csp_{ij}L_{\ssmQ \dsfa }^j)
\\   \nonumber
%\label{4line}
&=&-\frac{{\rm i}}{\sqrt{2}}(
\hat{L}^+L_\ssmS^{-i}\Csp_{ij} L_\ssmQ^{-j}
+\hat{L}^-L_\ssmQ^{+i}\Csp_{ij} L_\ssmS^{+j}
+\hat{L}^x L_\ssmS^{-i}\Csp_{ij} L_\ssmQ^{+j}
+\hat{L}^{\bar{x}} L_\ssmS^{+i}\Csp_{ij} L_\ssmQ^{-j})
\\
\label{5line}
&& +\ \frac{1}{\sqrt{2}}L_\ssmD (\frac{1}{2}L_\ssmS^{+i}\Csp_{ij} L_\ssmS^{-j}
+L_\ssmQ^{-i}\Csp_{ij} L_\ssmQ^{+j}) \ .
\end{eqnarray}
Eq.(\ref{3line})  provides
representation of the $AdS_5$  part  of the 3-form
  in the $sl(2)\oplus su(4)$ basis,
while (\ref{5line}) --  in the light-cone basis.

Let us now  consider  the
 $S^5$ part of the WZ  3-form in \rf{wzt}, i.e.
${\rm i}s^{IJ}L^\apr\bar{L}^I\gamma^\apr L^J$.
Representing  it
in terms of the  charged Cartan forms as in
(\ref{dec1}), \
${\cal H}_{S^5}={\cal H}_{S^5}^q+{\cal H}_{S^5}^{\bar{q}}$,
we get
\begin{eqnarray}
\label{1lines}
{\cal H}_{S^5}^q
&=&{\rm i} L^\apr
L^{q\alpha i}C_{\alpha\beta}(\Csp \gamma^\apr)_{ij} L^{q\beta j}
=-2L^{q\alpha i}C_{\alpha\beta}\Csp_{ik} L^k{}_j L^{q\beta j}
\\
\label{2lines}
&=&\frac{1}{2\sqrt{2}}L_\ssmS^{\sfa i}(\Csp L)_{ij}L_{\ssmS \sfa}^j
-\frac{1}{\sqrt{2}}L_\ssmQ^{\dsfa i}(\Csp L)_{ij}L_{\ssmQ \dsfa }^j
\\  \nonumber
%\label{3lines}
&=&\frac{1}{2\sqrt{2}}[L_\ssmS^{+i}(\Csp L)_{ij}L_\ssmS^{-j}
-L_\ssmS^{-i}(\Csp L)_{ij}L_\ssmS^{+j}]      \nonumber
\\
&& +\ \frac{1}{\sqrt{2}}[L_\ssmQ^{+i}(\Csp L)_{ij}L_\ssmQ^{-j}
-L_\ssmQ^{-i}(\Csp L)_{ij}L_\ssmQ^{+j}]   \ .
 \label{3lines}
\end{eqnarray}
Note that in (\ref{1lines}) we exploited  the relation (\ref{lijlab}) and used
the fact that $(\Csp \gamma^{\apr\bpr})_{ij}$ is symmetric in $i,j$,
the charge conjugation matrix 
$C_{\alpha\beta}$ is antisymmetric in $\alpha,\beta$ and    the
fermionic Cartan 1-forms $L^q$ are commuting with each other.
Eq.(\ref{2lines}) provides representation of $S^5$
part of the WZ 3-form
in the $sl(2)\oplus su(4)$ basis, while Eq.(\ref{3lines})
-- in the light-cone basis.

Next, let us  outline the procedure of derivation of the WZ term in
the light-cone $\kappa$-symmetry gauge.
Taking into account
that $L_\ssmS^{+i}=0$, $L_{\ssmS i}^+=0$, $L_\ssmK^+=0$,
$L_\ssmK^x=0$, $L_\ssmK^{\bar{x}}=0$
and plugging the Cartan
1-forms given by (\ref{carfor1})--(\ref{carfor2}) into the
 above expressions we get
${\cal H}_{AdS_5}^{q} ={\cal H}_{AdS_5}^{q(1)} +{\cal H}_{AdS_5}^{q(2)}$
where  (see \rf{tilthe},\rf{tildthe})
\begin{equation}
{\cal H}_{AdS_5}^{q(1)} =-\frac{{\rm i}}{\sqrt{2}}e^\phi
dx^+ \tilde{d\eta}^i\Csp_{ij} \tilde{d\theta}^j -\frac{{\rm
i}}{\sqrt{2}}e^\phi d\phi\tilde{d\theta}^i\Csp_{ij}\tilde{\eta}^jdx^+
\approx
d(\frac{{\rm i}}{\sqrt{2}}e^\phi dx^+\tilde{\eta}^i\Csp_{ij}\tilde{d\theta}^j)
\ , \end{equation}
\begin{equation}
{\cal H}_{AdS_5}^{q(2)}
=-\sqrt{2}e^\phi dx^+dx
\tilde{d\eta}^i\Csp_{ij}\tilde{\eta}^j
-\frac{1}{\sqrt{2}}e^\phi d\phi dx^+dx\tilde{\eta}^i\Csp_{ij}\tilde{\eta}^j
\approx
d(-\frac{1}{\sqrt{2}}e^\phi dx^+dx\tilde{\eta}^i\Csp_{ij}\tilde{\eta}^j)
\ . \end{equation}
The signs $\approx$ indicate  that these relations are valid
modulo terms  which are obtained by acting  by  differential $d$ on the
matrix
$U^i{}_j$ which enters in the definition of $\tilde{\eta}$,
 $\tilde{d\theta}$.
Such $d U^i{}_j$ terms are  canceled
 by contributions  coming from the
  $S^5$ part of WZ 3-form  which in  the light-cone gauge
takes the form
\begin{eqnarray}
{\cal H}_{S^5}^q
=\frac{1}{\sqrt{2}}[L_\ssmQ^{+i}(\Csp L)_{ij}L_\ssmQ^{-j}
-L_\ssmQ^{-i}(\Csp L)_{ij}L_\ssmQ^{+j}] \ .
\end{eqnarray}
To summarize,  one gets  the following exact relation
\be
{\cal H}_{AdS_5}^q+{\cal H}_{S^5}^q
=d\Bigl[\frac{{\rm i}}{\sqrt{2}}e^\phi dx^+\tilde{\eta}^i\Csp_{ij}
(\tilde{d\theta}^j+{\rm i}\tilde{\eta}^jdx)\Bigr]
\ . \ee
Multiplying this expression by ${\rm i}$, adding the Hermitean conjugate
and going from the 3-d  to the 2-d
 representation of the WZ term
 gives  the WZ  part of the string Lagrangian
${\cal L}_{WZ}$ in (\ref{actwz2}).

%%%%%%%%%%%%%%%%%%%%%%%%%%%%%%%%%%%%%%%%%%%%%%%%%%%%%%%%%%%
\appendix{\adss action in  S-gauge }
%%%%%%%%%%%%%%%%%%%%%%%%%%%%%%%%%%%%%%%%%%%%%%%%%%

The  results for the  Cartan forms in the
$sl(2)\oplus su(4)$ basis described
in Appendix B  allow us to  find  another  version of  the
$\kappa$-symmetry gauge fixed action of  superstring in $AdS_5\times S^5$.
Let us  start with the supercoset representative
(cf. \rf{kilgau}--\rf{kapfixG})
\begin{equation}\label{G2}
G= g_{x,\theta}\  g_\eta \ g_y \ g_\phi \ ,
\end{equation}
\be
 g_{x,\theta} =  \exp(x^aP^a +\theta_i^\sfa Q_\sfa^i
-\theta^{\dsfa i} Q_{{\dsfa }i}) \ ,\ee
\be
 g_\eta  =  \exp(\eta^{\sfa i}S_{\sfa i}-\eta_i^{\dsfa }S_{\dsfa }^i)
 \ , \ee
and impose the $\kappa$-symmetry gauge  by
\begin{equation}\label{sgau}
\eta^{\sfa i}=\eta_i^{\dsfa }=0 \ ,
\end{equation}
i.e.
\be
   G_{g.f.}=g_{x,\theta}\ g_y\ g_\phi \ .
   \ee
Since we have set to zero
 the fermionic coordinates  $\eta$ which correspond to
  the conformal supercharges $S$
 we  shall call this S-gauge.\footnote{The  ``S-gauge" and ``Q-gauge" terminology
was introduced in \ci{kallr},
but  our S-gauge is different from the one used in
\ci{kallr}.}
The resulting  gauge fixed  expressions for the
Cartan 1-forms are given by
\begin{eqnarray}
&&
L_P^a=e^\phi\Bigl[dx^a -\frac{\rm i}{2\sqrt{2}}(
\theta_{i\sfa}\sigma^{a\sfa \dsfb}d\theta_{\dsfb}^i
+\theta^{i\dsfa}\bar{\sigma}^a_{\dsfa\sfb}d\theta_i^{\sfb})\Bigr] \ ,
\la{PPP}\\
&&
L_{\ssmQ i}^\sfa =e^{\phi/2}\tilde{d\theta}_i^\sfa
\ , \ \ \ \ \
L_{\ssmQ}^{\dsfa i} =e^{\phi/2}\tilde{d\theta}^{\dsfa i}   \ ,
\\
&& \label{scarforij}
L_D=d\phi\ , \ \ \ \ \   \ \    L^i{}_j=(dUU^{-1})^i{}_j    \ ,
\end{eqnarray}
where $\tilde{d\theta}$ is  defined as in
(\ref{tildthe}) while the matrix $U$ is defined by
(\ref{uuu}),(\ref{dere}). All the remaining
 Cartan 1-forms are equal  to zero.

 Using that now   $L_{\ssmS}=0$ we get
from (\ref{3line}) the following expressions for the
$AdS_5$ part of the  3-form $  \cal H$
\be
{\cal H}_{AdS_5}^q=\frac{1}{2\sqrt{2}}d\phi e^{\phi}
\tilde{d\theta}^{\dsfa i}\Csp_{ij}\tilde{d\theta}_{\dsfa}^j  \ ,
\ee
while eq. (\ref{2lines}) gives
\be
{\cal H}_{S^5}^q=-\frac{1}{\sqrt{2}} e^{\phi}
\tilde{d\theta}^{\dsfa i}(\Csp L)_{ij}\tilde{d\theta}_{\dsfa}^j  \ .
\ee
Thus  we conclude that
 \be
{\cal H}_{AdS_5}^q+{\cal H}_{S^5}^q
=d\Bigl(\frac{1}{2\sqrt{2}}e^\phi
\tilde{d\theta}^{\dsfa i}\Csp_{ij}\tilde{d\theta}_{\dsfa}^j\Bigr) \ ,
\ee
which allows us to find the 2-d form of  the WZ term.

 Using  the above
relations and  (\ref{lkin1}),(\ref{hll}) and taking  into account
that $L_{\ssmK}^a=0$ we  finally
get the following kinetic and WZ parts of the \adss
string  Lagrangian  (cf. \rf{actkin2}--\rf{actwz2})
\be\label{c12}
{\cal L}_{kin}=-\frac{1}{2}\sqrt{g}g^{\vm\vn}
(L_{\ssmP \vm}^aL_{\ssmP \vn}^a
+\partial_\vm\phi\partial_\vn \phi+e_\vm^\apr e_\vn^\apr)
\ , \ee
\be
{\cal L}_{WZ}=\frac{\rm i}{2\sqrt{2}}\epsilon^{\vm\vn}
e^\phi \partial_\vm\theta^{\dsfa i}C_{ij}^U
\partial_\vn\theta_{\dsfa}^j +h.c.
\ ,  \ee
where $L_{\ssmP \vm}^a$ is given by   \rf{PPP} and $C_{ij}^U$ as in (\ref{CU}).
Note that in this S-gauge the 1-form $L^\apr$ which is given  in terms of
$L^i{}_j$ as in (\ref{hll}) is equal  simply  to the $S^5$   1-form
 $e^\apr$. The reason is that,  in contrast to what happens in  the
  light-cone gauge (\ref{carforij}), here
the Cartan form $L^i{}_j$ does not  contain
 fermionic contributions (see
(\ref{scarforij})). Making use of formula (\ref{intrel})
we get the following manifestly $SU(4)$ invariant representation for WZ part

\be
{\cal L}_{WZ}=\frac{\rm i}{2\sqrt{2}}\epsilon^{\vm\vn}
\partial_\vm\theta^{\dsfa i}\rho_{ij}^M Y^M
\partial_\vn\theta_{\dsfa}^j +h.c.
\ee
This form of WZ action by using usual $SO(6)$ $\gamma$ matrices (\ref{usgam}) 
can be cast
into the form similar to the one given in \ci{pessan,kalram} 
(see also \ci{KT}). The kinetic term (\ref{c12}) 
can be transformed into $SU(4)$ 
manifestly invariant form in a standard way.
Our presentation gives selfcontained derivation of $SU(4)$ manifestly 
invariant action from the original $5+5$ form of action given in \ci{MT}.

%%%%%%%%%%%%%%%%%%%%%%%%%%%%%%%

\end{document}